\documentclass[nofootinbib,aps,preprint,superscriptaddress]{revtex4}
\usepackage{epsfig}
\usepackage{color}

\newcommand{\fms}[1]{{#1}\!\!\!/}
\newcommand{\fmsl}[1]{{#1}\!\!\!\!/}
\newcommand{\half}{\frac{1}{2}}

\newcommand{\mP}{\mathcal{P}}
\newcommand{\bP}{\overline{\mathcal{P}}}
\newcommand{\Dp}{i\fmsl{D}^{\perp}}
\newcommand{\nn}{\frac{\fms{\overline{n}}}{2}} 
\newcommand{\N}{\frac{\fms{n}}{2}} 
\newcommand{\pn}{\frac{\fms{n}\fms{\overline{n}}}{4}} 
\newcommand{\pnb}{\frac{\fms{\overline{n}}\fms{n}}{4}} 

\newcommand{\n}{\overline{n}}
\newcommand{\bu}{\overline{u}}
\newcommand{\bs}{\overline{s}}

\newcommand{\w}{\omega}
\newcommand{\ka}{\kappa}
\newcommand{\LL}{\Biggl(\frac{\alpha_s (\mu)}{\alpha_s (\mu_0)}\Biggr)}

\newcommand{\B}{\mathcal{B}^{\perp}}
\newcommand{\p}{\overline{n} \cdot p}

\begin{document}

\vspace*{18pt}
\title{Leading SU(3) breaking in lightcone distribution amplitudes}

\baselineskip 3.0ex

\def\addDuke{Department of Physics, Duke University, Durham NC 27708, USA} 
\def\addPitt{Department of Physics and Astronomy, University of Pittsburgh, PA 15260, USA}
\author{Chul Kim}\email{chul@phy.duke.edu}\affiliation{\addDuke} 
\author{Adam K. Leibovich}\email{akl2@pitt.edu}\affiliation{\addPitt}


\begin{abstract} \vspace*{18pt}
\baselineskip 3.0ex 
The lightcone formalism including SU(3) breaking effects for the light pseudoscalar mesons is studied using soft-collinear effective theory (SCET), where the conformal symmetries needed for the expansion can be clearly implemented.  The lightcone distribution amplitudes (LCDAs) are well-defined at each order in the SCET power counting, $\lambda$.  Relations between the LCDAs are reproduced using the SCET formalism.  Treating the SU(3) breaking perturbatively,  the leading breaking effects can be described in a simple manner.  As a result, a new relationship among the LCDAs for the light mesons $\pi,~K$, and $\eta$ is developed, valid to leading order in the SU(3) breaking.

\end{abstract}

\maketitle

\section{Introduction}

Processes which include an energetic light meson in the initial or final state are often described by factoring the long distance interactions of the light meson into a lightcone distribution amplitude (LCDA).
The LCDA depends on the large momentum fractions of the partons inside the light meson.  Power counting is achieved using a twist expansion on the lightcone.  Due to the conformal symmetry on the lightcone (see e.g.~\cite{Braun:2003rp}), the form of the LCDA can be expanded in an infinite series of polynomials with definite conformal spin.  The coefficients of the series do not mix at one-loop order.  Assuming the series converge, the nonperturbative physics can be described by the first few coefficients, which can hopefully be extracted from experiment. 

Soft-collinear effective theory (SCET)  \cite{SCET1,SCET2} was developed to describe processes that include highly energetic collinear particles interacting with soft degrees of freedom, precisely the situation discussed above.  Factoring the collinear and soft degrees of freedom is achieved in SCET by a field redefinition.  To obtain the effective theory, first hard degrees of freedom with the offshellness $p^2 \sim Q^2$ are integrated out, and operators in full QCD are matched onto SCET systematically in powers of $\lambda\sim\Lambda_{\rm QCD}/Q$.  The Wilson coefficients for the SCET operators give the perturbative hard parts of the interaction, while the matrix elements of the SCET operators are nonperturbative functions describing the long distance interactions.  For example, at leading order in $\lambda$, the pion form factor $F_{\pi\gamma}$ is
\begin{eqnarray} 
F_{\pi\gamma} &=& \frac{i}{2Q} \int^1_0 du ~T_H (Q, u) ~\left\langle \pi^0 \left|\bar{\xi}_n W \delta\left(u-\frac{\bP^{\dagger}}{\n\cdot p_{\pi}}\right) \nn \gamma_5 \frac{\lambda^3}{\sqrt{2}} W^{\dagger} \xi_n \right| 0\right \rangle \nonumber \\
\label{fpg}
&=& f_{\pi} \int^1_0 du ~T_H(Q, u) ~\phi_{\pi} (u),
\end{eqnarray}  
where $\lambda^3$ is a Gell-Mann matrix with normalization $\mathrm{Tr}[\lambda^a \lambda^b] = 2 \delta^{ab}$, the lightcone vector $n^{\mu}$ and $\n^{\mu}$ satisfy $n^2=\n^2=0$ and $n\cdot \n =2$, and 
$\bP = \n\cdot \mP$ is a derivative operator giving the large label momentum in the $n$-direction from the collinear fields. $W$ is a collinear Wilson line defined as 
\begin{equation} 
\label{CW}
W (x) = \exp \Biggl( -g \frac{1}{\bP} \n\cdot A_{n} (x) \Biggr)
= \mathrm{P} \exp \Biggl( ig \int^x_{-\infty} ds \n\cdot A_n (s\n^{\mu})\Biggr),
\end{equation}
where P represents the path-ordered integral. 
In Eq.~(\ref{fpg}), we used the definition of the leading LCDA in SCET \cite{LCSCET} 
\begin{equation} 
-i f_{\pi} \frac{\n\cdot p_{\pi}}{2} \phi_{\pi}(u) = \langle \pi^0 |\bar{\xi}_n W \delta\Bigl(u-\frac{\bP^{\dagger}}{\n\cdot p_{\pi}}\Bigr) \nn \gamma_5 \frac{\lambda^3}{\sqrt{2}} W^{\dagger} \xi_n | 0 \rangle,
\end{equation} 
which in coordinate space is equivalent to
\begin{equation} 
\langle \pi^0 |\bar{\xi}_n W (x) \nn \gamma_5 \frac{\lambda^3}{\sqrt{2}} W^{\dagger} \xi_n (y) | 0 \rangle = -if_{\pi} \frac{\n\cdot p_{\pi}}{2} \int^1_0 du e^{i\n\cdot p_{\pi} (ux + \bar{u} y)} \phi_{\pi} (u). 
\end{equation} 
We have defined the momentum fraction $\bu = 1-u$, and the coordinate $x^{\mu}$ and $y^{\mu}$ are on the same lightcone such that $x^{\mu} = x\n^{\mu}, ~~y^{\mu} = y \n^{\mu}$ (this notation will be used throughout this paper). As can be seen from Eq.~(\ref{CW}), the multiplication of the collinear Wilson lines $W (x)$ and $W^{\dagger} (y)$ is simply interpreted as a Wilson link 
\begin{equation} 
[x, y] = W (x) W^{\dagger} (y) = \mathrm{P} \exp \Biggl( ig \int^x_{y} ds \n\cdot A_n (s\n^{\mu})\Biggr).
\end{equation} 
The matrix element of SCET is well-defined 
and can be compared to the leading twist LCDA in full QCD. 

Although LCDAs in full QCD are well-defined using the twist expansions, the formalism can be quite difficult to apply to higher order processes.
SCET may be a useful here since the factorization can be simply obtained.  Furthermore, there are useful constraints on the SCET operators using  the gauge symmetries \cite{SCET2} and reparameterization invariance (RPI) \cite{RPI} of SCET.
The remarkable point is that the twist expansion of the lightcone formalism in full QCD is equivalent to the power expansion of $\lambda$ in SCET. 
As investigated in Ref.~\cite{Hardmeier:2003ig}, it is possible to give one-to-one correspondence between the matrix elements in SCET and the lightcone formalism at any order in $\lambda$. 
Thus the matrix elements of subleading collinear operators in SCET automatically give the higher-twist LCDAs.  

Flavor SU(3) is a good symmetry for the hard kernels of factorized high-energy processes, because SU(3) breaking effects are usually suppressed by at least $1/Q$. 
However in nonperturbative functions, such as LCDAs for $K$ and $B_s$, the SU(3) breaking where  $m_s \gg m_{u,d}$ can be large and a main source of hadronic uncertainties \cite{Leibovich:2003jd}. With the rough estimate $m_s/\Lambda_{\rm QCD} \sim 0.3$, there are significant corrections to the SU(3) limit.  Furthermore, in the heavy quark limit ($m_Q \to \infty$) or the large energy limit ($E \to \infty$), these SU(3) breaking corrections do not vanish, so SU(3) corrections must be treated differently than the expansions in $1/m_Q$ or $1/E$.  Previous studies of SU(3) breaking effects on LCDA for light mesons have been considered in the framework of lightcone sum rule (LCSR) \cite{sumrule,Ball:2006wn} and chiral perturbation theory (ChPT) \cite{Chen:2003fp,Chen:2005js}.  The results are important to precise predictions of exclusive hadronic $B$ decays \cite{exclusive} and recently developed semi-inclusive hadronic $B$ decays \cite{Chay:2006ve}.

The SU(3) breaking effects can also be studied systematically using SCET.  
For convenience we keep only the strange quark mass ignoring $u$ and $d$ quark masses in the SCET Lagrangian. In this case the breaking effects come from the following quark mass terms in SCET Lagrangian 
\cite{Leibovich:2003jd}
\begin{equation} 
\label{mass}
\label{mterm}
\mathcal{L}_m^{(1)} = m ~\bar{\xi}_n \Biggl[\Dp,\frac{1}{\n\cdot iD} \Biggr] \nn \xi_n, ~~~
\mathcal{L}_m^{(2)} = -m^2 \bar{\xi}_n \frac{1}{\n\cdot iD} \nn \xi_n,
\end{equation} 
where the covariant derivative for the collinear field has been defined and power-counted as 
\begin{eqnarray} 
\label{dev}
iD^{\mu} &=& (\bP + g\n\cdot A_n) \frac{n^{\mu}}{2} + (\mP_{\perp}^{\mu} + gA^{\mu}_{n,\perp} ) 
+ (n\cdot \mP + g n\cdot A_n) \frac{\n^{\mu}}{2}  \\
&=& \mathcal{O} (\lambda^0) + \mathcal{O} (\lambda^1) + \mathcal{O} (\lambda^2). \nonumber 
\end{eqnarray} 
The mass terms in the Lagrangian originate from the following decomposition of the collinear quark field
\begin{equation} 
\psi (x) = \sum_{\tilde{p}} e^{-i\tilde{p}\cdot x} (\xi_n + \xi_{\n} ) 
= \sum_{\tilde{p}} e^{-i\tilde{p}\cdot x} \Biggl[ \xi_n + \frac{1}{\n\cdot iD}\Bigl(\Dp + m\Bigr)\nn \xi_n \Biggr],
\end{equation} 
where $\xi_n$ and $\xi_{\n}$ satisfy the constraints 
\begin{eqnarray} 
\frac{\fms{n}\fms{\n}}{4} \xi_n &=& \xi_n, ~~~\frac{\fms{\n}\fms{n}}{4} \xi_n = 0, \\
\frac{\fms{\n}\fms{n}}{4} \xi_{\n} &=& \xi_{\n}, ~~~\frac{\fms{n}\fms{\n}}{4} \xi_{\n} = 0.
\end{eqnarray} 
Applying the QCD equation of the motion, we can relate $\xi_{\n}$ to $\xi_n$, 
\begin{equation} 
\label{xinb}
\xi_{\n} = \frac{1}{\n\cdot iD} \Bigl(i\fmsl{D}^{\perp}+m \Bigr) \nn 
\xi_n,
\end{equation} 
where we see that $\xi_{\n}$ is suppressed by $\mathcal{O} (\lambda)$ compared to $\xi_n$. 
SU(3) breaking effects occur via the time-ordered products of $\mathcal{L}_m^{(1),(2)}$ or the $\xi_{\n}$ components in SCET operators at the higher order in $\lambda$.  With the scaling $E \gg \Lambda \gg m_s$, SCET allows for two independent power series expansions, $\lambda$ and $m_s/\Lambda$. At each order in $\lambda$ we can describe the leading SU(3) breaking effects in LCDAs, which can be compared to previous works. 

In this paper, we derive relations between the LCDAs of $\pi$, $K$, and $\eta$ which cover the leading SU(3) breaking effects. At leading order in $\lambda$, the relation can be easily derived through the time-ordered products of $\mathcal{L}_m^{(1)}$ and twist-2 operators. Further, it can be extended to twist-3 LCDAs and can be confirmed by the exact relations between twist-3 LCDAs, $\phi_p^M$, $\phi_{\sigma}^M$, and $\phi_{3M}$. The rest of the paper is organized as follows.  In Sec. II, we briefly discuss conformal symmetry in SCET. In Sec. III, the lightcone formalism for energetic pseudoscalar mesons is introduced in the framework of SCET.  Using isospin symmetry in case of the pion, we derive constraints between the twist-3 LCDAs equivalent to the established results in full QCD \cite{Braun:1989iv}. In Sec. IV and V, we investigate the leading and subleading SU(3) breaking effects, respectively, using SCET. In Sec. VI, using constraints between LCDAs, we obtain exact forms of the twist-3 LCDAs $\phi_p^M$ and $\phi_{\sigma}^M$ (or $\phi_{\pm}^M$). With SU(3) broken, those LCDAs are expressed in terms of coefficients of the Gegenbauer polynomials in $\phi_M$ together with nonperturtative parameters in $\phi_{3M}$. Finally, we conclude in Sec. VII.

\section{Conformal symmetry in SCET}
Lightcone conformal symmetry for QCD~\cite{Braun:2003rp} is extremely useful for the analysis of nonlocal operators in SCET.  In general, SCET fields are 
considered to have definite conformal spin $j$ and twist $t$ 
\begin{equation} 
j=\half (l+s), ~~~t=l-s, 
\end{equation} 
where $l$ denotes dimension of the field and 
$s$ is a collinear spin obtained by applying the collinear generator
$\mathcal{S}_n$  to the field 
\begin{equation} 
\mathcal{S}_n \Phi = \half \n^{\mu}n^{\nu} \Sigma_{\mu\nu}  \Phi  
= s\Phi.
\end{equation}   
The spin operator $\Sigma_{\mu\nu}$ depends on the representation of fields, 
\begin{equation} 
\Sigma_{\mu\nu} \phi = 0,~~\Sigma_{\mu\nu} \psi = \frac{i}{2} \sigma_{\mu\nu} \psi,~~
\Sigma_{\mu\nu} A_{\alpha}  = g_{\nu\alpha} A_{\mu} - g_{\mu\alpha} A_{\nu}.
\end{equation}  
The collinear quark fields in SCET have the following collinear 
spins 
\begin{equation} 
\mathcal{S}_n \xi_n = \half \Bigl(\pn - \pnb \Bigr) \xi_n = \half \xi_n,~~~
\mathcal{S}_n \xi_{\n} = \half \Bigl(\pn - \pnb \Bigr) \xi_{\n} = -\half 
\xi_{\n}.
\end{equation} 
The gauge invariant SCET fields 
have their own conformal spins and twists as follows \cite{Hill:2004if}:   
\begin{eqnarray} 
&&W^{\dagger}\xi_n (x) ~:~ 
j=1,~~t=1 \nonumber \\
&&W^{\dagger}\xi_{\n} (x) = \frac{1}{\bP} W^{\dagger} \Bigl( 
i\fmsl{D}^{\perp} + m \Bigr) \nn \xi_n (x)~:~ j=\half,~~ t=2 \nonumber \\
&&\Bigl[\bP W^{\dagger} iD^{\perp\mu} W \Bigr](x) ~:~ j=\frac{3}{2},~~t=1, \\
&&\Bigl[\bP W^{\dagger} n\cdot iD W \Bigr](x) ~:~ j=1,~~t=2. 
\nonumber 
\end{eqnarray} 
In SCET power-counting, $\xi_n \sim \mathcal{O} (\lambda)$ and each component of $iD^{\mu}$ is power-counted as shown in Eq.~(\ref{dev}). Therefore we find that the twist expansion is identical with the $\lambda$ expansion.    

In general the nonlocal operator which consists of two fields on the lightcone with definite conformal spins $j_1$ and $j_2$,
\begin{equation}
\mathcal{O} (\alpha_1,\alpha_2) = \Phi_{j_1} (\alpha_1\n) \Gamma 
\Phi_{j_2} (\alpha_2\n),
\end{equation}
can be constructed by the conformal towers of the highest-weight local operators with $j=j_1+j_2+n$,
\begin{equation} 
\mathcal{O}_n^{j_1,j_2}(\alpha) = (\n\cdot i\partial)^n
\Biggl[ \Phi(\alpha \n) \Gamma P_n^{(2j_1-1,2j_2-1)}
\Biggl(\frac{\n\cdot i\partial-\n\cdot i\overleftarrow{\partial}}
{\n\cdot i\partial+\n\cdot i\overleftarrow{\partial}}\Biggr)
\Phi(\alpha \n)\Biggr],
\label{conop}
\end{equation}   
where $P_n^{(\alpha,\beta)}$ are the Jacobi Polynomials.
The conformal local operators at each $n$ in Eq.~(\ref{conop}) do 
not mix under renormalization to leading order because the 
renormalization group equation (RGE) is no more than the Ward identity for 
the dilatation generator of the conformal group \cite{Braun:2003rp}. 

As an example, consider the leading order (twist-2) light cone wave function for pion, 
\begin{equation} 
\langle \pi^+ | \bar{\xi}_n^u W (x) \nn \gamma_5 W^{\dagger}\xi_n^d (y) | 0 
\rangle = -if_{\pi} \frac{\n\cdot p_\pi}{2} \int^1_0 du e^{i\n\cdot p_{\pi} (ux+\bu y)}
\phi_{\pi} (u).
\label{phi}
\end{equation}
The nonlocal operator on the light cone can be expanded in terms of 
\begin{equation} 
\mathcal{O}_n^{1,1}(x) 
= \bar{\xi}^u_n W (x) \nn \gamma_5 (\bP_-)^n
C_n^{3/2} \Biggl( \frac{\bP_+}{\bP_-}\Biggr) 
W^{\dagger} \xi_n^d (x), 
\end{equation}  
where $\bP_{\pm} = \bP^{\dagger} \pm \bP$ and the Gegenbauer polynomials $C_n^{3/2} \sim P_n^{(1,1)}$. 
When we take the vacuum-to-pion matrix element, the result is 
\begin{equation} 
\label{con}
\langle \pi^+ | \mathcal{O}_n^{1,1}(x)  | 0 \rangle = -\frac{i}{2} f_{\pi}
(\n\cdot p_{\pi})^{n+1} \int^1_0 du C_n^{3/2} (2u-1) \phi_{\pi} (u,\mu).  
\end{equation}  
As discussed above, due to the conformal symmetry the above equation does not mix under renormalization with the terms $m \neq n$.  Futher,  there is no mixing with  
collinear gluon operators with $j=3+n-1$ due to the flavor structure of the pion.\footnote{In the flavor singlet case, it would mix with the two-gluon operator.  The SCET calculation of the RGE mixing for the twist-2 LCDAs has been performed in \cite{Fleming:2004rk}.} 

In order to take advantage of the orthogonality of $C_n^{3/2}$,
\begin{equation} 
\label{org}
\int^1_0 du u\bu C_n^{3/2} (2u-1) C_m^{3/2}(2u-1) = \delta_{mn} 
\frac{(n+1)(n+2)}{4(2n+3)}, 
\end{equation} 
we expand $\phi_{\pi}(u,\mu)$ in Gaugenbauer polynomials  
\begin{equation} 
\phi_{\pi}(u,\mu) = 6u\bu \sum_{n=0} a_n (\mu) C_n^{3/2}(2u-1),
\label{phiform}
\end{equation} 
where $a_0 = 1$ from the normalization condition $\int du\, \phi_{\pi} (u) =1 $.
Then, from Eq.~(\ref{con}), the coefficient $a_n (\mu)$ are multiplicatively renormalized \cite{Lepage:1980fj,Chernyak:1983ej}, 
\begin{equation}  
\label{anren} 
a_n (\mu) = a_n (\mu_0) \Biggl(\frac{\alpha_s (\mu)}{\alpha_s(\mu_0)} \Biggr)^{\gamma_n^{(0)}/\beta_0},
\end{equation} 
where $\beta_0 = 11-2n_f/3$ is the first coefficient of the QCD $\beta$ function, and the anomalous dimension for $a_n$ is   
\begin{equation} 
\label{gamman}
\gamma_n = \frac{\alpha_s C_F}{2\pi} \gamma_n^{(0)}= \frac{\alpha_s C_F}{2\pi} \Biggl[1-\frac{2}{(n+1)(n+2)} + 4 \sum_{k=2}^{n+1} \frac{1}{k}\Biggr].
\end{equation} 
Since $\gamma_n$ with $n \neq 0$ is positive definite, the coefficients for the higher conformal spins can be neglected when the renormalization scale $\mu\gg\mu_0$.

\section{LCDAs for Pseudoscalar mesons} 

In full QCD, the LCDAs for pseudoscalar mesons to twist-3 accuracy are defined as 
\begin{eqnarray} 
\label{fphi}
\langle M | \bar{q} (x) \gamma_{\mu} \gamma_5 [x,y] q(y) | 0 \rangle 
&=& -i f_{M} \n\cdot p_M \frac{n_{\mu}}{2} 
 \int^1_0 du e^{i\n\cdot p_M (ux+\bu y)} \phi_{M} (u), \\
\label{phip}
\langle M | \bar{q} (x) \gamma_5 [x,y] q(y) | 0 \rangle 
&=& -i f_{M} \mu_{M} \int^1_0 du e^{i\n\cdot p_M (ux+\bu y)} \phi_p^M(u),   \\
\langle M | \bar{q} (x) \sigma^{\mu\nu} \gamma_5 [x,y] q(y) | 0 \rangle 
\label{phis}
&=& i f_{M} \frac{\mu_{M}}{6} (p_M^{\mu}\Delta^{\nu} - p_M^{\nu}\Delta^{\mu}) \nonumber\\
&&\times \int^1_0 du e^{i\n\cdot p_M (ux+\bu y)} \phi_{\sigma}^M (u),   \\
\langle M | \bar{q} (x) \sigma^{\mu\nu} \gamma_5 [x,z] gG^{\alpha\beta} (z) [z,y] 
q(y) | 0 \rangle &=& i[p_M^{\mu}(p_M^{\alpha}g^{\nu\beta}-p_M^{\beta}g^{\alpha\nu})
-(\mu \leftrightarrow \nu) ] f_{3M} \nonumber \\  
&&\times \int^1_0 [du_i] 
e^{i\n\cdot p_M (u_1 x + u_2 y + u_3 z)} \phi_{3M}(u_i), 
\end{eqnarray} 
where $\Delta_{\mu} = x_{\mu} - y_{\mu} = (x-y) \n_{\mu}$ and $[du_i] = du_1du_2du_3 \delta (1-u_1-u_2-u_3)$. 

In SCET, we define the light cone wave function keeping the exact 
$\lambda~(\sim p_{\perp}/\n\cdot p \sim \Lambda /E_{M})$ power 
counting with definite conformal spins and twists as 
\begin{eqnarray} 
\label{phim}
\langle M | \bar{\xi}_n W (x) \nn \gamma_5 W^{\dagger}\xi_n (y) | 0 
\rangle &=& -if_{M} \frac{\n\cdot p_M}{2} \int^1_0 du e^{i\n\cdot p_M (ux+\bu y)}
\phi_{M} (u),  \\ 
\label{phi+}
\langle M | \bar{\xi}_n W (x) \gamma_5 W^{\dagger}\xi_{\n} (y) | 0 
\rangle &=& -if_{M}\frac{\mu_{M}}{2} 
\int^1_0 du e^{i\n\cdot p_M(ux+\bu y)} \phi_+^M (u), \\
\label{phi-}
\langle M | \bar{\xi}_{\n} W (x) \gamma_5 W^{\dagger}\xi_{n} (y) | 0 
\rangle &=& -if_{M}\frac{\mu_{M}}{2} 
\int^1_0 du e^{i\n\cdot p_M (ux+\bu y)} \phi_-^M (u), \\ 
\label{gm}
\langle M | \bar{\xi}_{\n} W (x) \N \gamma_5 W^{\dagger}\xi_{\n} (y) 
| 0 \rangle &=& -if_{M} \frac{n\cdot p_M}{2} 
\int^1_0 du e^{i\n\cdot p_M (ux+\bu y)} g_{M} (u), \\
\label{3M} 
\langle M | \bar{\xi}_{n} W (x) \nn \gamma^{\perp}_{\mu} \B_{\nu} (z) 
 \gamma_5 W^{\dagger}\xi_{n} (y) |0 \rangle
&=& \frac{i}{2} f_{3M} g^{\perp}_{\mu\nu} (\n\cdot p_M)^2 \nonumber \\
&&\times \int^1_0 [du_i] e^{i\n\cdot p_M(u_1x+u_2y+u_3z)} \phi_{3M} (u_i),  
\end{eqnarray}
where $\B_{\nu}(z) = \Bigl[\bP W^{\dagger} iD_{\nu}^{\perp} W \Bigr] (z)$. 
Eqs.~(\ref{phi+}), (\ref{phi-}), and (\ref{3M}) are $\lambda$-suppressed 
(twist-3) compared to the leading order, while Eq.~(\ref{gm}) is 
$\lambda^2$-suppressed. 

To include the effects of a light quark mass, we need the SCET Lagrangian including the quark mass  \cite{Leibovich:2003jd},
\begin{eqnarray} 
\mathcal{L}_{\mathrm{SCET}} &=& \bar{\xi}_n n\cdot iD \nn \xi_{n} 
+\bar{\xi}_n \Bigl(i\fmsl{D}_{\perp} - m\Bigr) \xi_{\n} 
+\bar{\xi}_{\n} \Bigl(i\fmsl{D}_{\perp} - m\Bigr) \xi_{n} 
+\bar{\xi}_{\n} \n\cdot iD \N \xi_{\n} \nonumber \\
\label{L}
&=&\bar{\xi}_n \Biggl[ n\cdot iD + 
i\fmsl{D}_{\perp} \frac{1}{\n\cdot iD_c} 
i\fmsl{D}_{\perp} \Biggr] \nn \xi_n + \mathcal{L}_m^{(1)} + \mathcal{L}_m^{(2)},
\end{eqnarray} 
where $\mathcal{L}_m^{(1),(2)}$ are given in Eq.~(\ref{mterm}), and for the second equality 
we used the relation in Eq.~(\ref{xinb}).      
Keeping the lowest order in $\lambda$ in Eqs.~(\ref{phip}) and (\ref{phis}),
we find the relations   
\begin{equation} 
\phi_p (u) = \half [\phi_+ (u) + \phi_- (u) ],~~~
\frac{\partial}{\partial u} \phi_{\sigma} (u) = 
 -3 [\phi_+ (u) - \phi_- (u) ].
\label{3rel}
\end{equation}   
Furthermore, expanding Eq.~(\ref{fphi}) to twist-4 accuracy gives \cite{Ball:1998je}
\begin{eqnarray}   
\langle M | \bar{q} (x) \gamma_{\mu} \gamma_5 [x,y] q(y) | 0 \rangle 
&=& -i f_{M} \n\cdot p_M \frac{n_{\mu}}{2} 
 \int^1_0 du e^{i\n\cdot p_M (ux+\bu y)} \phi_{M} (u) \nonumber \\
\label{fgpi}
&&-\frac{i}{2} f_{M} m_{M}^2 \frac{1}{p_M\cdot \Delta}\Delta_{\mu} 
 \int^1_0 du e^{i\n\cdot p_M (ux+\bu y)}g_{M} (u),
\end{eqnarray} 
where the second term on the right-hand side corresponds to Eq.~(\ref{gm}).

The coefficient $\mu_{M}$ in Eqs.~(\ref{phip}) and (\ref{phis}), which results from quark condensation, can be obtained by the equation of motion in full QCD. For the pion, consider a translation of 
Eq.~(\ref{fphi}) in the limit $y \to x =0$, 
\begin{eqnarray}  
P^{\mu} \langle \pi^+ | \bar{u} \gamma_{\mu} \gamma_5 d(0) | 0 
\rangle 
&=& -if_{\pi} p_{\pi}^2 = -if_{\pi} m_{\pi}^2 \nonumber \\
&=& 
\langle \pi^+ | \bar{u} \Bigl(-i\overleftarrow{\fms{\partial}}+i\fms{\partial}
\Bigr)  \gamma_5  d(0) | 0 \rangle =
\langle \pi^+ | \bar{u} \Bigl(i\overleftarrow{\fmsl{D}}+ i\fmsl{D}\Bigr)
 \gamma_5  d(0) | 0 \rangle \nonumber \\ 
&=& (m_u + m_d) 
\langle \pi^+ | \bar{u} \gamma_5 d(0) | 0 \rangle, 
\end{eqnarray} 
where $i\overleftarrow{D}_{\mu} = -i\overleftarrow{\partial}_{\mu} + 
g A_{\mu}$.  We then see, using Eq.~(\ref{phip}), $\mu_\pi = m_{\pi}^2/(m_u+m_d)$. 

In SCET the quark condensation factor $\mu_M$ can also be acquired easily. 
We begin by decomposing the pion momentum as 
\begin{equation} 
p^{\mu} = \n\cdot p \frac{n^{\mu}}{2} + p_{\perp}^{\mu} 
+ n\cdot p \frac{\n^{\mu}}{2} = \n\cdot p \frac{n^{\mu}}{2} + 
 \frac{m_{\pi}^2}{\n\cdot p} \frac{\n^{\mu}}{2}.
\label{pion}
\end{equation} 
Taking the limit $x\to y = 0$ in Eqs.~(\ref{phim}) and (\ref{gm}), we obtain 
\begin{eqnarray} 
\label{effrel}
-if_{\pi} m_{\pi}^2 &=& -if_{\pi} (\n\cdot p n\cdot p) 
= -if_{\pi} (\half \n\cdot p n\cdot p + \half n\cdot p \n\cdot p)  \\
&=& n\cdot P \langle \pi^+ | \bar{\xi}_n^u \nn \gamma_5 \xi_n^d 
(0) | 0 \rangle +
\n \cdot P 
\langle \pi^+ | \bar{\xi}_{\n}^u  \N \gamma_5 \xi_{\n}^d (0) 
| 0 \rangle \nonumber \\
&=& \langle \pi^+ | \bar{\xi}_n^u  \nn \Bigl(n\cdot \mathcal{P}^{\dagger}
-n\cdot \mathcal{P} \Bigr) \gamma_5 \xi_n^d | 0 \rangle 
+ \langle \pi^+ | \bar{\xi}_{\n}^u  \N \Bigl( \bP^{\dagger}
- \bP \Bigr) \gamma_5 \xi_{\n}^d | 0 \rangle 
\nonumber \\
&=& \langle \pi^+ | \bar{\xi}_n^u  \nn \Bigl(n\cdot i\overleftarrow{D}
-n\cdot iD \Bigr) \gamma_5 \xi_n^d  | 0 \rangle 
+ \langle \pi^+ | \bar{\xi}_{\n}^u  \N \Bigl(\n\cdot i\overleftarrow{D}
-\n\cdot iD \Bigr) \gamma_5 \xi_{\n}^d | 0 \rangle, \nonumber 
\end{eqnarray}
where $i\overleftarrow{D}_{\mu}$ is given by 
$\mathcal{P}^{\dagger}_{\mu}+ gA_{\mu}$.
From the SCET equation of motion and using Eq.~(\ref{xinb}), 
the second and the third bilinear operators in the last equality lead to 
\begin{eqnarray}  
\bar{\xi}_{n}^u  \gamma_5
n\cdot iD \nn \xi_{n}^d &=&
-\bar{\xi}_{n}^u  \gamma_5
\Bigl(i\fmsl{D}_{\perp}-m_d \Bigr) \frac{1}{\n\cdot iD} 
\Bigl(i\fmsl{D}_{\perp}+m_d \Bigr) \nn \xi_n^d,\\
\bar{\xi}_{\n}^u  \N \n\cdot i\overleftarrow{D}
\gamma_5 \xi_{\n}^d  &=& \bar{\xi}_{n}^u  
\Bigl(-i\overleftarrow{\fmsl{D}_{\perp}}+m_u \Bigr) \gamma_5
\frac{1}{\n\cdot iD} 
\Bigl(i\fmsl{D}_{\perp}+m_d \Bigr) \nn \xi_n^d \nonumber \\
&=& \bar{\xi}_{n}^u  \gamma_5
\Bigl(i\fmsl{D}_{\perp}+m_u \Bigr) \frac{1}{\n\cdot iD} 
\Bigl(i\fmsl{D}_{\perp}+m_d \Bigr) \nn \xi_n^d, 
\end{eqnarray}
where we used $i\overleftarrow{D}_{\perp}^{\mu} = 
iD_{\perp}^{\mu}$ since 
we chose a frame where the total transverse momentum of the partons in the pion system are  
zero, Eq.~(\ref{pion}). Combining these equations, we 
find 
\begin{equation} 
\bar{\xi}_{n}^u \gamma_5
n\cdot iD \nn \xi_{n}^d + 
\bar{\xi}_{\n}^u  \N \n\cdot i\overleftarrow{D}
\gamma_5 \xi_{\n}^d  =
(m_u + m_d) \bar{\xi}_{n}^u  \gamma_5 \frac{1}{\n\cdot iD} 
\Bigl(i\fmsl{D}^{\perp}+m_d \Bigr) \nn \xi_n^d. 
\end{equation} 
Similarly, the first and fourth operators in the last equality of Eq.~(\ref{effrel}) lead to 
\begin{equation} 
\bar{\xi}_{n}^u  \nn n\cdot i\overleftarrow{D}\gamma_5 \xi_{n}^d +   
\bar{\xi}_{\n}^u  \gamma_5 \n\cdot iD \N \xi_{\n}^d = 
(m_u + m_d) \bar{\xi}_{n}^u  \nn 
\Bigl(i\overleftarrow{\fmsl{D}_{\perp}}+m_u \Bigr) 
\frac{1}{\n\cdot i\overleftarrow{D}} \gamma_5 \xi_{n}^d. 
\end{equation} 
Finally, we find, up to SU(2) corrections,
\begin{eqnarray} 
-i f_{\pi} m_{\pi}^2 &=& (m_u+m_d) \langle \pi^+ | 
\bar{\xi}_{n}^u  \gamma_5 \frac{1}{\n\cdot iD} 
\Bigl(i\fmsl{D}_{\perp}+m_d \Bigr) \nn \xi_n^d +
\bar{\xi}_{n}^u  \nn 
\Bigl(i\overleftarrow{\fmsl{D}_{\perp}}+m_u \Bigr) 
\frac{1}{\n\cdot i\overleftarrow{D}} \gamma_5 \xi_{n}^d | 0 \rangle 
\nonumber \\
&\sim& 2(m_u+m_d)\langle \pi^+ | 
\bar{\xi}_{n}^u  \gamma_5 \frac{1}{\n\cdot iD_c} 
\Bigl(i\fmsl{D}^{\perp}+m_d \Bigr) \nn \xi_n^d | 0 \rangle \\
&\sim& 2(m_u+m_d) \langle \pi^+ | \bar{\xi}_{n}^u  \nn 
\Bigl(i\overleftarrow{\fmsl{D}^{\perp}}+m_u \Bigr) 
\frac{1}{\n\cdot i\overleftarrow{D}} \gamma_5 \xi_{n}^d | 0 \rangle. \nonumber
\end{eqnarray} 
We can therefore obtain the chiral condensation factor $\mu_{\pi}$ in a well-defined manner  
in SCET. It can give a numerically sizable correction to subleading order 
matrix elements in the effective theory even though $\mu_M$ is power-counted as $\mathcal{O} (\lambda)$. 
In SU(3) limit, we identify $\mu_K \sim \mu_{\pi} \sim B_0$, where $B_0$ is proportional to the vacuum expectation value of the light quark pair,
\begin{equation} 
B_0 \propto \langle \bar{u} u \rangle \sim \langle \bar{d} d \rangle \sim \langle \bar{s} s \rangle. 
\end{equation} 
However including SU(3) breaking with nonzero $m_s$, we treat $\mu_K \neq \mu_{\pi}$ since $\langle \bar{s} s \rangle$ can be different from $\langle \bar{q} q \rangle$.  
For the kaon, the exact value for $\mu_K$ is $\mu_K=m_K^2/(m_s +m_q)\sim B_0 + B_1$, where $q=u, d$, and $B_1$ is the leading SU(3) breaking term from the vacuum expectation value of the light quark pairs. 
In order to investigate leading SU(3) breaking effects in LCDAs, it is useful to keep only the leading SU(3) breaking correction.  
We will therefore identify  $\mu_K = \mu_{\pi} + \mathcal{O} (m_s/\Lambda)$ in our notation without any further assumption. 

\subsection{Expansion of the subleading LCDAs using conformal symmetry} 

Because the SCET operators at each order have definite conformal spins, 
it is useful to expand these nonlocal operators in terms of local conformal operators.  
In this subsection, for simplicity we neglect the quark mass.
For the two-particle (or two-point) nonlocal operators,   
we can easily construct the conformal operators using Eq.~(\ref{conop}).   
The relevant conformal operators for the higher-order, nonlocal operators in 
Eqs.~(\ref{phi+}-\ref{gm}) are
\begin{eqnarray} 
\bar{\xi}_n^u W (x)  \gamma_5 W^{\dagger}\xi_{\n}^d (y)
&\longrightarrow& \mathcal{O}_n^{1,1/2} =  
\bar{\xi}_n^u W   \gamma_5 (\bP_-)^n P_n^{(0,1)} \Biggl(
\frac{\bP_+}{\bP_-}\Biggr)  W^{\dagger}\xi_{\n}^d, \\ 
\bar{\xi}_{\n}^u W (x) \gamma_5 W^{\dagger}\xi_{n}^d (y)
&\longrightarrow& \mathcal{O}_n^{1/2,1} = \bar{\xi}_{\n}^u W  \gamma_5 (\bP_-)^n P_n^{(1,0)} \Biggl(
\frac{\bP_+}{\bP_-}\Biggr)  W^{\dagger}\xi_n^d, \\ 
\bar{\xi}_{\n}^u W (x) \N \gamma_5 W^{\dagger}\xi_{\n}^d (y)
&\longrightarrow& \mathcal{O}_n^{1/2,1/2} = \bar{\xi}_{\n}^u W \N \gamma_5 (\bP_-)^n C_n^{1/2} \Biggl(
\frac{\bP_+}{\bP_-}\Biggr) W^{\dagger}\xi_{\n}^d,
\end{eqnarray} 
where we used the Jacobi polynomial identity  
$P_n^{(\alpha,\beta)} (-x) = (-1)^n 
P_n^{(\beta,\alpha)} (x)$, and in the last equation we
replaced the $P_n^{(0,0)}$ with the Gegenbauer polynomial $C_n^{1/2}$.
At one loop order, these operators for each $n$ do not mix. 
From the orthogonality relations 
\begin{eqnarray}
\int^1_0 du (u\bu)^{l-1/2} C_m^l (2u-1) C_n^l (2u-1) 
&=& \frac{\pi 2^{1-4l} \Gamma(2l+n)}{n!(n+l)\Gamma^2(l)}\delta_{mn},\\
\int^1_0du  \bu^{\alpha} u^{\beta} P_m^{(\alpha,\beta)}(2u-1)  
P_n^{(\alpha,\beta)}(2u-1) &=& \frac{\Gamma(\alpha+n+1)\Gamma(\beta+n+1)}
{n! (\alpha+\beta+1+2n) \Gamma(\alpha+\beta+n+1)}\delta_{mn},\nonumber
\end{eqnarray} 
we can determine the forms of the  LCDAs $\phi_+^{\pi},~\phi_-^{\pi}$, and $g_{\pi}$,
\begin{eqnarray}
\label{+}
\phi_+^{\pi} (u) &=& 2u\sum_{n=0} b^+_{n}(\mu) P_n^{(0,1)}(2u-1), \\ 
\label{-}
\phi_-^{\pi}(u) &=& 2\bu\sum_{n=0} b^-_{n}(\mu) P_n^{(1,0)}(2u-1), \\ 
g_{\pi}(u) &=& \sum_{n=0} c_{n}(\mu) C_n^{1/2}(2u-1). 
\end{eqnarray}

The three particle wave function $\phi_{3\pi}$ is more complicated. 
Because the relevant nonlocal operator can be expanded in terms of the local 
operatos with the conformal spin $j=7/2+n$,   
the explicit form can be written as \cite{Braun:2003rp,Braun:1989iv}
\begin{eqnarray} 
\phi_{3\pi}(u_i,\mu) &=& 360u_1u_2u_3^2\Bigl[\w^{7/2}(\mu) 
+\w^{9/2}(\mu)\half(7u_3-3) +\w^{11/2}_1(\mu)(2-4u_1u_2-8u_3+8u_3^2) 
\nonumber \\
\label{3pi}
&&\phantom{360u_1u_2u_3^2\Bigl[}+\w_2^{11/2}(\mu)(3u_1u_2-2u_3+3u_3^2) + \cdots\Bigr],
\end{eqnarray}
where $u_i = u_1, u_2, u_3$.  The coefficients $\w^{7/2}$ and $\w^{9/2}$ are multiplicatively  renormalized at one loop, while there is mixing between 
$\w^{11/2}_1$ and $\w^{11/2}_2$ with the anomalous dimensions given 
in Ref~\cite{Braun:1989iv}. 

\subsection{Relations between the subleading LCDAs}

At subleading order, there are constraints relating the three LCDAs $\phi_+^{\pi},~\phi_-^{\pi}$,
and $\phi_{3\pi}$ to each other. 
Using operator identities \cite{Balitsky:1987bk,Ball:1998ff}
based on the equation of motion,  
relations between the twist-3 LCDAs were introduced in Refs.~\cite{Ball:1998je,Braun:1989iv}.  
In SCET, the equivalent constraints can be obtained by analyzing the relevant operators directly. 
                                                                         
We start from the following identity 
\begin{equation} 
\langle \pi^+ | \bar{\xi}_n^u W (x) \nn \gamma_5 \Bigl(
\fmsl{\mathcal{P}}_{\perp}^{~\dagger}
- \fmsl{\mathcal{P}}_{\perp} \Bigr) W^{\dagger} \xi_n^d (y) |0 \rangle = 0, 
\label{st}
\end{equation} 
due to the fact that we can pick a frame where the 
total transverse momentum of the pion is zero.   
Using the identity 
\begin{equation} 
\label{iden1} 
\mathcal{P}^{\mu} W^{\dagger} \xi_n (x) 
= W^{\dagger} iD^{\mu}  \xi_n (x) + i \int^x_{-\infty} dz \Bigl[\bP W^{\dagger} iD^{\mu} 
W \Bigr] (z) W^{\dagger} \xi_n (x), 
\end{equation} 
where $[W^{\dagger} iD^{\mu} W] (x) = -i \int^x_{-\infty} dz [\bP W^{\dagger} iD^{\mu} 
W] (z)$, the bilinear operator in Eq.~(\ref{st}) can be written as  
\begin{eqnarray}  
\bar{\xi}_n^u W (x) \nn \gamma_5 \Bigl(
\fmsl{\mathcal{P}}_{\perp}^{~\dagger}
- \fmsl{\mathcal{P}}_{\perp} \Bigr) W^{\dagger} \xi_n^d (y) &=&
-\bar{\xi}_n^u \nn \overleftarrow{\Dp} W(x) \gamma_5 W^{\dagger} \xi_n^d (y) 
+\bar{\xi}_n^u \nn W(x) \gamma_5 W^{\dagger} \Dp  \xi_n^d (y)   \nonumber \\
&&-i\bar{\xi}_n^u W(x) \nn \gamma_5 \int^x_y dz \Bigl[\bP W^{\dagger}\Dp
W \Bigr] (z) W^{\dagger} \xi_n^d (y). 
\end{eqnarray}
Taking the matrix element of the above equation neglecting the quark mass, we obtain 
\begin{eqnarray} 
&&\left(-\frac{if_{\pi}\mu_{\pi}}{2}\right)
i\partial_x \int^1_0 du e^{i\p_{\pi} (ux+\bu y)} \phi_-^{\pi} (u)
-\Bigl(-\frac{if_{\pi}\mu_{\pi}}{2}\Bigr)
i\partial_y \int^1_0 du e^{i\p_{\pi} (ux+\bu y)} \phi_+^{\pi} (u)  \\
&&\phantom{\Bigl(-\frac{if_{\pi}\mu_{\pi}}{2}\Bigr)
i\partial_x}+(\p_{\pi})^2 f_{3\pi} \int^x_y dz \int [du_i] e^{i\p(u_1x +u_2y+u_3z)} 
\phi_{3\pi} (u_i) \nonumber \\
&&\quad= \frac{f_{\pi} \mu_{\pi}}{2} i\p_{\pi} \int^1_0 du e^{i\p_{\pi} (ux+\bu y)} 
\left[ u\phi_-^{\pi} (u)-\bu\phi_+^{\pi} (u) \right] \nonumber \\ 
&&\qquad-i\p_{\pi} f_{3\pi} \int [du_i] \left(e^{i\p_{\pi} [(u_1+u_3)x + u_2 y]}
-e^{i\p_{\pi} [(u_1x + (u_2+u_3) y]}\right) \frac{1}{u_3} \phi_{3\pi}(u_1,u_2,u_3)=0.\nonumber
\end{eqnarray}
From the second equality of the above equation, we find 
\begin{eqnarray}
&&R_{\pi} \int^1_0 du e^{i\p_{\pi} (ux+\bu y)} \Bigl[ \bu\phi_+^{\pi} (u) - u\phi_-^{\pi} (u)\Bigr] 
\nonumber \\
\label{relation}
&&\qquad= 2 \int^1_0 du e^{i\p_{\pi}(ux+\bu y)} 
\Biggl[\int^{\bu}_0 db \frac{\phi_{3\pi} (u,\bu-b,b)}{b} 
-\int^{u}_0 db \frac{\phi_{3\pi} (u-b,\bu,b)}{b} \Biggr],
\end{eqnarray}  
where $R_{\pi}=f_{\pi}\mu_{\pi}/f_{3\pi}$. 
Finally, due to isospin symmetry,
there is the following constraint 
\begin{equation}
\label{isorel} 
\phi_+^{\pi} (u) = \phi_-^{\pi} (\bu),\qquad~~~\phi_-^{\pi} (u) = \phi_+^{\pi} (\bu).
\end{equation}  

From these constraints, we can relate  
the coefficients in $\phi_{\pm}^{\pi}$ to combinations of the coefficients of $\phi_{3\pi}$. 
To begin with, as in Eqs.~(\ref{+}) and (\ref{-}), we expand $R_{\pi} \phi_{\pm}^{\pi}$ as
\begin{eqnarray} 
\label{R+}
R_{\pi}\phi_+^{\pi} (u) &=& 2u\Bigl[\kappa^{3/2} + \kappa^{5/2} P_1^{(0,1)}(2u-1)
+ \kappa^{7/2} P_2^{(0,1)}(2u-1)+ \kappa^{9/2} P_3^{(0,1)}(2u-1) 
\nonumber \\
&&\phantom{2u\Bigl[}+ \kappa^{11/2} P_4^{(0,1)}(2u-1)+\cdots \Bigr], \\  
R_{\pi} \phi_-^{\pi} (u) &=& 2\bu\Bigl[\kappa^{3/2} - \kappa^{5/2} P_1^{(1,0)}(2u-1)
+ \kappa^{7/2} P_2^{(1,0)}(2u-1) - \kappa^{9/2} P_3^{(1,0)}(2u-1) 
\nonumber \\
\label{R-}
&&\phantom{2u\Bigl[}+ \kappa^{11/2} P_4^{(1,0)}(2u-1)+\cdots \Bigr],  
\end{eqnarray} 
which satisfy Eq.~(\ref{isorel}), due to the fact $P_n^{(0,1)}(2u-1)=(-1)^n P_n^{(1,0)}(2\bu-1)$.
Then, substituting Eqs.~(\ref{3pi}), (\ref{R+}), and (\ref{R-}) into 
Eq.~(\ref{relation}), we find 
\begin{eqnarray} 
&&\kappa^{3/2} = R_{\pi},~~~\kappa^{5/2} = 0,~~~\kappa^{7/2} = 30\w^{7/2}, \nonumber
\\ &&\kappa^{9/2} = -3\w^{9/2},~~~\kappa^{11/2} = \frac{3}{2}
\Bigl(4\w^{11/2}_1 - \w^{11/2}_2\Bigr). 
\end{eqnarray}  
We can also determine $\phi_p^{\pi}$ and $\phi_{\sigma}^{\pi}$ from Eq.~(\ref{3rel}). 
Using the following identities 
\begin{eqnarray}
&&uP_n^{(0,1)}(2u-1)+\bu P_n^{(1,0)}(2u-1) = C_n^{1/2}(2u-1), \nonumber \\
&&uP_n^{(0,1)}(2u-1)-\bu P_n^{(1,0)}(2u-1) = C_{n+1}^{1/2}(2u-1), \\
&&\frac{d}{du}\Bigl[u\bu C_n^{3/2}(2u-1)\Bigr] 
= -\half (n+1)(n+2)C_{n+1}^{1/2}(2u-1), \nonumber 
\end{eqnarray}
we obtain 
\begin{eqnarray}
R_{\pi}\phi_p^{\pi} &=&R_{\pi}  + 30\w^{7/2} C_2^{1/2}(2u-1) +\frac{3}{2}\Bigl(
4\w^{11/2}_1-\w^{11/2}_2
-2\w^{9/2}\Bigr)C_4^{1/2}(2u-1)+\cdots, \\
R_{\pi} \phi_{\sigma}^{\pi} &=& 6u\bu \Bigl[
R_{\pi} + \Bigl(5\w^{7/2}-\half\w^{9/2}\Bigr) C_2^{3/2}(2u-1) \nonumber \\ 
&&~~~~~~
+\frac{1}{10}\Bigl(4\w^{11/2}_1-\w^{11/2}_2\Bigr)C_4^{3/2}(2u-1)+\cdots 
\Bigr]. 
\end{eqnarray}

\section{SU(3) breaking effects at leading order in $\lambda$}
\begin{figure}[t]
\begin{center}
\vspace{0.6cm}
\epsfig{file=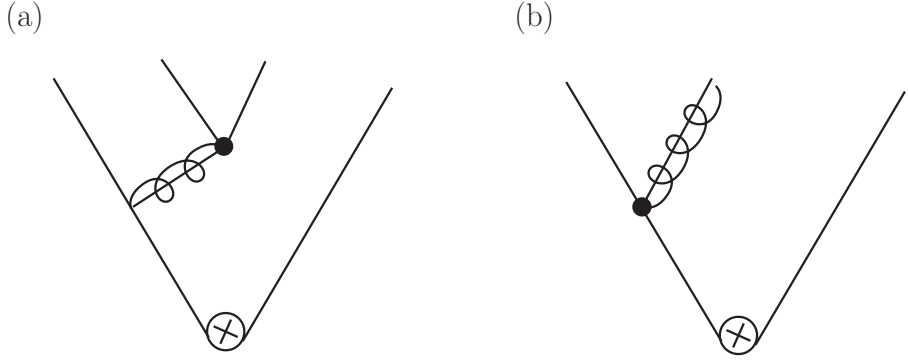, width=12cm}
\end{center}
\caption{\baselineskip 3.0ex  Leading SU(3) breaking interactions in LCDAs. The diagram (a,b) are examples of $\mathcal{T}_m^{(S,V)}$, respectively. The dot represents $\mathcal{L}_{m_s}^{(1)}$ and `$\otimes$' denotes the nonlocal SCET operator.  } 
\label{masstime}
\end{figure}

The large strange quark mass, compared to the up and down quarks, leads to significant SU(3) breaking.  We will only keep the strange quark mass in the analysis of SU(3) breaking, setting the $u$ and $d$ quark masses to zero.  Furthermore, we will use the approximation that $m_s/\Lambda$ is small, as was done in, e.g., Ref.~\cite{Leibovich:2003jd}.
In SCET, the leading SU(3) breaking effects for the LCDA are obtained from time-ordered products of $\mathcal{L}_{m_s}^{(1)}$ and the nonlocal twist-2 SCET operator,
\begin{equation} 
\label{timeo} 
\langle M | \mathrm{T} \Biggl\{ i\int d^4 z \mathcal{L}_{m_s}^{(1)} (z), \bar{\xi}_n^q W \delta \Bigl(x - \frac{\bP^{\dagger}}{\n\cdot p_M}\Bigr) \nn \gamma_5 W^{\dagger} \xi_n^{q'} \Biggr\}| 0 \rangle = \mathcal{T}_m^{(S)}(x, \mu) + \mathcal{T}_m^{(V)} (x, \mu),
\end{equation}
where $\{q,q'\} = \{u,d,s\}$, and $\mathcal{L}_{m_s}^{(1)}\sim \mathcal{O} (m_s/\Lambda)$ is given in Eq.~(\ref{mterm}) and can be rewritten as 
\begin{equation} 
\mathcal{L}_{m_s}^{(1)} = - m_s \bar{\xi}_n^s W \Biggl[\frac{1}{\bP} W^{\dagger} i\fmsl{D}_{\perp} W\Biggr] \nn W^{\dagger} \xi_n^s. 
\end{equation} 
Here the collinear derivative operator acts only within the square brackets, and thus $\mathcal{L}_{m_s}^{(1)}$ must involve at least one collinear gluon field.  This helps us  categorize the long distance contributions for the time-ordered products as so-called ``sea'' contributions $\mathcal{T}_m^{(S)}$ and ``valence'' contributions $\mathcal{T}_m^{(V)}$ \cite{Chen:2003fp}, where $\mathcal{L}_{m_s}^{(1)}$ are mediated by a collinear gluon ($\mathcal{T}_m^{(S)}$) and quark ($\mathcal{T}_m^{(V)}$) as shown in Fig.~\ref{masstime}. 

An important point is that the sea contribution is common to all the light mesons, independent of the  quark flavors of the nonlocal operators.  So the sea quark contribution cannot mediate SU(3) breaking effects between, for example, the pion and kaon. However the valence contribution only occurs for mesons with the strange valence quark component. Therefore we conclude that SU(3) breaking effects  can be specified by the valence contribution for each meson and we write the matrix element for the LCDA as  
\begin{equation} 
\label{general2}
\langle M | \bar{\xi}_n^q W \delta \Bigl(x - \frac{\bP^{\dagger}}{\n\cdot p_M}\Bigr) \nn \gamma_5 W^{\dagger} \xi_n^{q'} | 0 \rangle = \langle M | \bar{\xi}_n^q W \delta \Bigl(x - \frac{\bP^{\dagger}}{\n\cdot p_M}\Bigr) \nn \gamma_5 W^{\dagger} \xi_n^{q'} | 0 \rangle_{\mathrm{SU}(3)} 
+ \mathcal{T}_m^{(V)} (x, \mu),
\end{equation} 
where the subscript `SU(3)' represents the matrix element in the SU(3) limit, including the sea contribution.  It can be identified with the matrix element for pion,
\begin{equation} 
\label{su3a}
\langle M | \bar{\xi}_n^q W \delta \Bigl(x - \frac{\bP^{\dagger}}{\n\cdot p_M}\Bigr) \nn \gamma_5 W^{\dagger} \xi_n^{q'} | 0 \rangle_{\mathrm{SU}(3)} \sim -i \frac{f_{\pi}}{2} \n\cdot p_M \phi_{\pi} (x). 
\end{equation} 

The valence contribution in Eq.~(\ref{general2}) can be further specified by the quark flavors of the nonlocal operators, $\{\mathcal{T}_{m,A}^{(V)},~\mathcal{T}_{m,B}^{(V)},~ \mathcal{T}_{m,A+B}^{(V)}\} = \{(q=s~\text{and}~q'=u,d),~(q=u,d~\text{and}~q'=s),~(q=s~\text{and}~q'=s) \}$. 
The matrix element $\mathcal{T}_{m,A}^{(V)}$ contributes to $(K^-,\bar{K}^0)$ mesons, $\mathcal{T}_{m,B}^{(V)}$ to $(K^+,K^0)$ mesons, and $\mathcal{T}_{m,A+B}^{(V)}$ is needed for the $|s\bar{s} \rangle$ component of the $\eta$ meson. 
To leading order in the SU(3) breaking, we identify 
\begin{equation} 
\label{valence} 
\mathcal{T}_{m,A+B}^{(V)} (x) = \mathcal{T}_{m,A}^{(V)} (x) + \mathcal{T}_{m,B}^{(V)} (x),~~~
\mathcal{T}_{m,A}^{(V)} (x) = \mathcal{T}_{m,B}^{(V)} (1-x),
\end{equation} 
where the second equation follows from the charge conjugate symmetry of the strong interaction. 

\begin{figure}[t]
\begin{center}
\vspace{0.6cm}
\epsfig{file=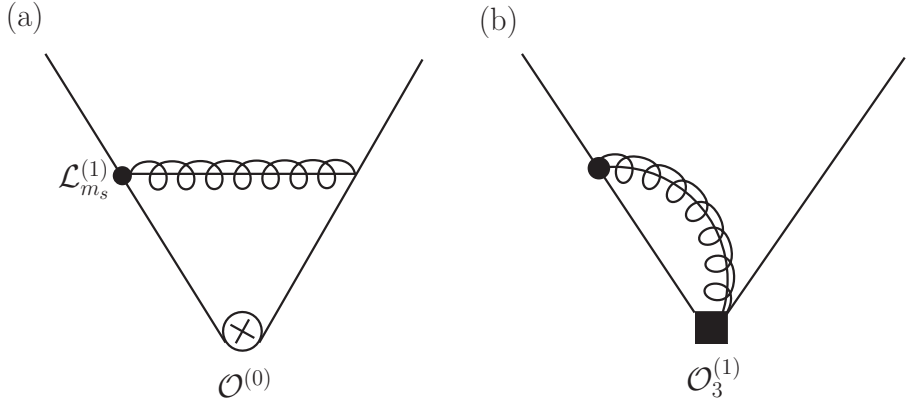, width=12cm}
\end{center}
\caption{\baselineskip 3.0ex  Examples of nontrivial renormalization effects for the time-ordered products of nonlocal SCET operators and the leading mass interaction term $\mathcal{L}_{m_s}^{(1)}$.  $\mathcal{O}^{(0)}$ in the diagram (a) denotes a leading twist operators shown in Eq.~(\ref{phim}), and $\mathcal{O}_3^{(1)}$ in the diagram (b) is a twist-3 3-particle operator which has been defined in Eq.~(\ref{3M}).  } 
\label{conmix}
\end{figure}

To find the distribution for $\mathcal{T}_{m}^{(V)}$, we need to consider the renormalization behavior of the time-ordered products in Eq.~(\ref{timeo}). The nontrivial renormalization can be calculated from diagrams such as Fig.~\ref{conmix}-(a), and it turns out to be zero because there is no mixing to other nonlocal operators at leading order in $\lambda$. We thus find that the time-ordered products can be renormalized as $\mathrm{T}\{Z_m^{(1)} \mathcal{L}_{m_s}^{(1)}, Z^{(0)} \otimes \mathcal{O}^{(0)} \}$, where $\mathcal{O}^{(0)}$ is a leading twist operator in the momentum space, 
and $\otimes$ denotes a convolution of the momentum fraction. As shown in Ref.~\cite{Chay:2005ck}, $\mathcal{L}_{m_s}^{(1)}$ is unrenormalized and so $Z_m^{(1)} = 1$.  Then the renormalization behavior of the time-ordered product becomes the same as the leading twist operator. As a result,  we can expand $\mathcal{T}_{m}^{(V)}$ in terms of Gegenbauer polynomials $C_n^{3/2}$,
\begin{eqnarray} 
\label{su3m}
\mathcal{T}_{m,\{A,B,A+B\}}^{(V)} (x,\mu) &=& -\frac{i}{2} f_{m,\{A,B,A+B\}} \n\cdot p_M  \phi_m^{\{A,B,A+B\}} (x,\mu) \\ 
&=&-\frac{i}{2} f_{m,\{A,B,A+B\}} \n\cdot p_M ~6x(1-x) \sum_{n=0} a_n^{m,\{A,B,A+B\}} (\mu) C_n^{3/2}(2x-1). \nonumber 
\end{eqnarray} 
For example in case of $K^-$, combining the above with with Eqs.~(\ref{phim}), (\ref{general2}), and (\ref{su3a}), we have    
\begin{equation} 
\label{kpi}
f_K \phi_{K^-} (x,\mu) = f_{\pi} \phi_{\pi} (x, \mu) + f_{m,A} \phi_{m}^A (x, \mu).
\end{equation} 

In Eq.~(\ref{su3m}), we have introduced a new, nonperturbative decay constant, $f_m$.  In order to specify the value of $f_m$, we need to decide the normalization for $\mathcal{T}_{m}^{(V)}$, with the choices $\int dx\, \phi_{m} (x) = 0$ or 1. If we choose the normalization as 0, we identify $f_K = f_{\pi}$ from Eq.~(\ref{kpi}). In this case the decay constants do not break SU(3) at this order,  and the breaking enters in the distribution of the light meson with the strange quark component. 
A more general choice is to assign to $f_K$ leading SU(3) breaking effects by setting the normalization of the valence quark distribution as $\int dx\, \phi_m (x) = 1$. In this case, we have the SU(3) breaking $\Delta_f \equiv f_K - f_{\pi} = f_{m,A} \sim \mathcal{O} (m_s)$ from Eq.~(\ref{kpi}). Experimentally the decay constants for pion and kaon are $f_{\pi} = 130.7 ~\mathrm{MeV},~f_{K} = 159.8~\mathrm{MeV}$ \cite{Eidelman:2004wy} with the ratio $f_K/f_{\pi} = 1.22$.
So there is significant SU(3) breaking in the decay constants, which prefers the normalization $\int dx\, \phi_m (x) =1$. 
Therefore, in our analysis, we have the SU(3) breaking correction
\begin{equation} 
\label{fk}
\frac{f_K}{f_{\pi}} = 1 + \frac{\Delta_f}{f_{\pi}} + \mathcal{O}\Bigl(\frac{m_s^2}{\Lambda^2}\Bigr), 
\end{equation}
where $\Delta_f / f_{\pi} = 0.22 + \mathcal{O} (m_s^2/\Lambda^2)$ numerically.

Identifying $\Delta_f = f_{m,A} = f_{m,B}$, we have the relation for $K^+$ 
\begin{equation} 
\label{kplus}
f_{K} \phi_{K^+} (x, \mu) = f_{\pi} \phi_{\pi} (x, \mu) + \Delta_f \phi_{m}^B (x, \mu), 
\end{equation} 
where $\phi_{m}^B(x) = \phi_{m}^A (1-x)$ from the second equation in Eq.~(\ref{valence}). 
The leading-twist matrix element for the $\eta$ meson,
\begin{equation} 
\langle \eta |\bar{\xi}_n W \delta\Bigl(x-\frac{\bP^{\dagger}}{\n\cdot p_{\eta}}\Bigr) \nn \gamma_5 \frac{\lambda^8}{\sqrt{2}} W^{\dagger} \xi_n | 0 \rangle = -i f_{\eta} \frac{\n\cdot p_{\eta}}{2} \phi_{\eta}(x,\mu), 
\end{equation} 
gives, by combining Eqs.~(\ref{general2}),~(\ref{valence}), and (\ref{su3m}),
\begin{equation} 
\label{eta} 
f_{\eta} \phi_{\eta} (x, \mu) = f_{\pi} \phi_{\pi} (x, \mu) + \frac{2}{3} \Delta_f \Bigl[ \phi_{m}^A  (x, \mu) + \phi_{m}^B  (x, \mu) \Bigr]. 
\end{equation} 
Here, we neglect $\eta-\eta'$ mixing, so the $\eta$ state is given by $|\eta\rangle = |\eta_8\rangle = (1/\sqrt{6})(|u\bar{u}\rangle+|d\bar{d}\rangle-2|s\bar{s}\rangle)$. Integrating both sides of Eq.~(\ref{eta}) over $x$ results in the relation 
\begin{equation} 
\label{feta}
f_{\eta} = f_{\pi} + \frac{4}{3} \Delta_f \sim f_{\pi} \Bigl(\frac{f_K}{f_{\pi}}\Bigr)^{4/3}, 
\end{equation} 
which is similar to the ChPT results \cite{Gasser:1984gg}.  If we now use isospin symmetry, Eqs.~(\ref{kpi}), (\ref{kplus}), and (\ref{eta}) lead to 
\begin{equation} 
\label{su3rel} 
f_{\pi} \phi_{\pi} (x, \mu) + 3f_{\eta} \phi_{\eta} (x, \mu) = 2f_K \Bigl[\phi_{K^-} (x, \mu) + \phi_{K^+} (x, \mu) \Bigr] = 2f_K \Bigl[\phi_{\bar{K}^0} (x, \mu) + \phi_{K^0} (x, \mu) \Bigr],
\end{equation} 
valid at leading order in the SU(3) breaking.

Since we can expand the LCDAs in terms of Gegenbauer polynomials, we can use Eq.~(\ref{su3rel}) to relate the coefficients of the various LCDAs,
\begin{equation} 
\label{an}
f_{\pi} a_{2n}^{\pi} (\mu) + 3f_{\eta} a_{2n}^{\eta} (\mu) = 4 f_K a_{2n}^{K} (\mu), ~~~n=0,1,2,\cdots, 
\end{equation} 
where all the zeroth coefficients $a_0^M =1$ due to the normalization of the LCDAs.  For $\phi_{\pi}$ and $\phi_{\eta}$, the coefficients with odd $n=2m+1$ vanish since the distributions are be invariant (up to isospin violations) under the replacement $x \leftrightarrow \bar{x}=1-x$ and $C_n (2x-1) = (-1)^n C_n (2\bar{x}-1)$. 
On the other hand, for $\phi_K$, the odd numbered coefficients $a_{2n+1}^K$ need not vanish due to SU(3) breaking between the quark and antiquark field.  We do have some relationships between coefficients, however.  From Eq.~(\ref{valence}), we have $a_n^{\{K^-,\bar{K}^0\}} = (-1)^n a_n^{\{K^+,K^0\}}$.
Furthermore, $a_{m,n}^A = (-1)^n a_{m,n}^B$ and $a_{m,n}^{A+B} = a_{m,n}^A + a_{m,n}^B$ in the valence distribution $\phi_m(x)$. Combining all the above leads to
\begin{eqnarray} 
\label{ak1su3} 
f_{\pi} a_{2n+1}^K &=& \Delta_f (a_{m,2n+1}^A-a_{2n+1}^K), \\
\label{ak2su3}
f_{\pi} (a_{2n}^K - a_{2n}^{\pi} ) &=& \Delta_f (a_{2n}^A - a_{2n}^K), \\ 
\label{aetasu3} 
f_{\pi} (a_{2n}^{\eta} - a_{2n}^{\pi} ) &=& \frac{4}{3} \Delta_f (a_{2n}^A - a_{2n}^{\eta}),
\end{eqnarray} 
where $a_n^K = a_n^{\{K^-,\bar{K}^0\}}$, and the decay constants $f_K$ and $f_{\eta}$ have been expanded in terms of $f_{\pi}$ using Eqs.~(\ref{fk}) and (\ref{feta}). 

The magnitude of the valence contribution is $\mathcal{O} (m_s) \sim \Delta_f$, which implies $a_{m,n} \sim \mathcal{O}(1)$.  If we turn off the SU(3) breaking corrections, $a_n^{\{K,\eta\}}$ reduces to  $a_n^{\pi}$, which suggests $a_{2n+1}^K \sim \mathcal{O} (m_s/\Lambda)$ and $a_{2n}^{\{K,\eta\}} \sim a_{2n}^{\pi} + \mathcal{O} (m_s/\Lambda)$. 
Therefore, up to the first order correction to SU(3), the above equations simplify to 
\begin{eqnarray} 
\label{ak1su3t} 
f_{\pi} a_{2n+1}^K &=& \Delta_f a_{m,2n+1}^A, \\
\label{ak2su3t}
f_{\pi} (a_{2n}^K - a_{2n}^{\pi} ) &=& \Delta_f (a_{2n}^A - a_{2n}^{\pi}), \\ 
\label{aetasu3t} 
f_{\pi} (a_{2n}^{\eta} - a_{2n}^{\pi} ) &=& \frac{4}{3} \Delta_f (a_{2n}^A - a_{2n}^{\pi}).
\end{eqnarray} 

\section{SU(3) breaking corrections at subleading order in $\lambda$}
The same analysis done at leading order in $\lambda$ can be applied to the matrix elements for LCDAs at subleading order.  Consider the following matrix elements for $K^-$ 
\begin{equation} 
\label{subpm}
\langle K | \mathcal{O}^{(1)}_{\pm} (x,\mu) | 0 \rangle = -\frac{i}{2} f_K \mu_K \phi_{\pm}^{K} (x,\mu), 
\end{equation} 
where $\mathcal{O}^{(1)}_{\pm}(x)$ are twist-3 nonlocal operators in momentum space
\begin{eqnarray} 
\label{o+1}
\mathcal{O}_{+}^{(1)} (x) &=& \mathcal{J}_{+} (x) + \mathcal{K}_{+} (x) = \bar{\xi}_n^q W \delta\Bigl(x-\frac{\bP^{\dagger}}{\n\cdot p_K}\Bigr)\frac{1}{\bP} \gamma_5 W^{\dagger}\Bigl( i\fmsl{D}_{\perp}+m_{q'} \Bigr) \nn  \xi_n^{q'}, \\
\label{o-1} 
\mathcal{O}_{-}^{(1)} (x) &=& \mathcal{J}_{-} (x) + \mathcal{K}_{-} (x) = \bar{\xi}_n^q \nn  \Bigl( \overleftarrow{i\fmsl{D}_{\perp}}+m_q \Bigr) W \frac{1}{\bP^{\dagger}}\delta\Bigl(x-\frac{\bP^{\dagger}}{\n\cdot p_K}\Bigr) \gamma_5 W^{\dagger} \xi_n^{q'}.
\end{eqnarray} 
Here, $\mathcal{K}_{\pm} (x)$ are the pieces of the above operators proportional to $m_q$ and $m_{q'}$, respectively and are suppressed by $\mathcal{O}(m_q/\Lambda)$ relative to $\mathcal{J}_{\pm}$. The quark and antiquark  flavors $q,~q'$ are chosen according to the flavor contents of the meson. For example, if we consider the matrix elements for $K^-$, the flavors $(q,q')$ are given by $(s,u)$, and $\mathcal{K}_+$ can be neglected since it is proportional to $m_u$. 
In Eq.~(\ref{subpm}), the leading SU(3) breaking corrections come from the two sources: the valence contribution from the time-ordered products of $\mathcal{L}_{m_s}^{(1)}$ and $\mathcal{J}_{\pm}$, and the matrix elements of $\mathcal{K}_{\pm}$.   

For the $K^-$ system, the matrix elements of Eq.~(\ref{subpm}) can be written as 
\begin{eqnarray} 
\label{k-+}
\langle K^- | \mathcal{O}_+^{(1)} (x) | 0 \rangle  &=& \langle K^- | \mathcal{J}_+ (x) |0  \rangle_{\mathrm{SU(3)}} + \mathcal{T}_{m,A}^{(V,+)} (x) \\ 
\label{k--}
\langle K^- | \mathcal{O}_-^{(1)} (x) | 0 \rangle &=& \langle K^- | \mathcal{J}_- (x) | 0 \rangle_{\mathrm{SU(3)}} + \langle K^- | \mathcal{K}_- (x) | 0 \rangle + \mathcal{T}_{m,A}^{(V,-)} (x), 
\end{eqnarray}
where $\mathcal{T}_{m,A}^{V,\pm}$ are the valence contribution of $\mathcal{J}_{\pm}$ for $K^-$.  The matrix elements with the subscript `SU(3)' are analogous to the treatment in Eq.~(\ref{su3a}) and can be expressed in terms of the pion LCDAs 
\begin{equation} 
\langle K | \mathcal{O}^{(1)}_{\pm} (x,\mu) | 0 \rangle_{\mathrm{SU(3)}} = -\frac{i}{2} f_{\pi} \mu_{\pi} \phi_{\pm}^{\pi} (x,\mu).
\end{equation} 
Similarly, the matrix elements for $K^+$ system can be written as 
\begin{eqnarray} 
\label{k++}
\langle K^+ | \mathcal{O}_+^{(1)} (x) | 0 \rangle &=& \langle K^+ | \mathcal{J}_+ (x) | 0 \rangle_{\mathrm{SU(3)}} + \langle K^+ | \mathcal{K}_+ (x) | 0 \rangle + \mathcal{T}_{m,B}^{(V,+)} (x), \\
\label{k+-}
\langle K^+ | \mathcal{O}_-^{(1)} (x) | 0 \rangle  &=& \langle K^- | \mathcal{J}_- (x) |0  \rangle_{\mathrm{SU(3)}} + \mathcal{T}_{m,B}^{(V,-)} (x). 
\end{eqnarray}
Due to charge conjugate symmetry, we have the relations  
\begin{equation} 
\mathcal{T}_{m,A}^{(V,\pm)} (x) = \mathcal{T}_{m,B}^{(V,\mp)} (\bar{x}),~~~
\langle K^- | \mathcal{K}_- (x) | 0 \rangle = \langle K^+ | \mathcal{K}_+ (\bar{x}) | 0 \rangle. 
\end{equation} 
Hence we identify $\phi_{\pm}^{K^-} (x) = \phi_{\mp}^{K^+} (\bar{x})$.  The same analysis is applicable to $\{\bar{K}^0,~K^0\}$.  

Next, consider the $\eta$ meson.  In this case, the valence contribution for the $|s\bar{s}\rangle$ state in $\eta$ can be expressed as $\mathcal{T}_{m,A+B}^{(V,\pm)} (x) = \mathcal{T}_{m,A}^{(V,\pm)} (x) + \mathcal{T}_{m,B}^{(V,\pm)} (x)$, giving the matrix elements 
\begin{equation} 
\label{eta3}
\langle \eta | \mathcal{O}_{\pm}^{(1)} (x) | 0 \rangle = \langle \eta | \mathcal{J}_{\pm} (x) | 0 \rangle_{\mathrm{SU(3)}} + \frac{2}{3} \langle \eta | \mathcal{K}_{\pm} (x) | 0 \rangle + \frac{2}{3} \Bigl(\mathcal{T}_{m,A}^{(V,\pm)} (x) + \mathcal{T}_{m,B}^{(V,\pm)} (x)\Bigr).
\end{equation}
Finally, combining the above equations, we find 
\begin{equation} 
f_{\pi} \mu_{\pi} \phi_{\pm}^{\pi} (x) + 3 f_{\eta} \mu_{\eta} \phi_{\pm}^{\eta} (x) = 2f_K \mu_{K} 
\Bigl[\phi_{\pm}^{K^-} (x) + \phi_{\pm}^{K^+} (x) \Bigr] = 2f_K \mu_{K} 
\Bigl[\phi_{\pm}^{\bar{K}^0} (x) + \phi_{\pm}^{K^0} (x) \Bigr],  
\label{pmrel}
\end{equation} 
or, using Eq.~(\ref{3rel}),
\begin{equation} 
f_{\pi} \mu_{\pi} \phi_{p,\sigma}^{\pi} (x) + 3 f_{\eta} \mu_{\eta} \phi_{p,\sigma}^{\eta} (x) = 2f_K \mu_{K} 
\Bigl[\phi_{p,\sigma}^{K^-} (x) + \phi_{p,\sigma}^{K^+} (x) \Bigr] = 2f_K \mu_{K} 
\Bigl[\phi_{p,\sigma}^{\bar{K}^0} (x) + \phi_{p,\sigma}^{K^0} (x) \Bigr].
\label{psrel}
\end{equation} 
Choosing the normalization $\int dx\, \phi_{\pm,p,\sigma}^{M} (x) =1$, we have 
\begin{equation} 
\label{fmurel} 
f_{\pi} \mu_{\pi} + 3 f_{\eta} \mu_{\eta}  = 4 f_K \mu_{K}, 
\end{equation} 
which, using $f_K = f_{\pi} + \Delta_f,~f_{\eta} = f_{\pi} + 4\Delta_f/3$, can be rewritten as 
\begin{equation} 
\label{fmurel2} 
f_{\pi} (\mu_{\pi} + 3\mu_{\eta} -4\mu_K) = 4 \Delta_f (\mu_K - \mu_{\eta} ).
\end{equation} 
If we take the lowest order in SU(3) breaking, the right side of the above equation should vanish since $\mu_{K,\eta} = \mu_{\pi} + \mathcal{O} (m_s/\Lambda)$. This implies
\begin{equation} 
\label{murel} 
\mu_{\pi} +3\mu_{\eta} = 4\mu_K. 
\end{equation} 

The contributions $\mathcal{T}_{m}^{(V,\pm)} (x)$ cannot be easily determined because the renormalization of the relevant time-ordered product mixes with a time-ordered product involving $m_s \mathcal{O}^{(0)}$. As a result, $\phi_{\pm}^{K,\eta} (x)$ involves leading-twist Gegenbauer coefficients $a_n^{K,\eta}$.  As seen in Section III, $\phi_{\pm}^{M}$ can be related to the three-particle state LCDAs $\phi_{3M}$.  Including SU(3) breaking, the relation is modified, as we shall see, but a useful relation is still available.
We can express $\phi_{\pm}^{K,\eta}$ in terms of $a_n^{K,\eta}$ and coefficients of $\phi_{3K,3\eta}$, which will be investigated in the next section.  

Applying a similar analysis to the three-particle state LCDAs for $K,~\eta$, the matrix elements, to  first order in SU(3) breaking,  can be written as 
\begin{eqnarray} 
\langle K^- | \mathcal{O}_3^{(1)} (x_i,\mu) | 0 \rangle &=& \langle K^- | \mathcal{O}_3^{(1)} (x_i,\mu) | 0 \rangle_{\mathrm{SU(3)}} + \mathcal{T}_{3m,A}^{(V)} (x_i), \nonumber \\ 
\label{3su3} 
\langle K^+ | \mathcal{O}_3^{(1)} (x_i,\mu) | 0 \rangle &=& \langle K^+ | \mathcal{O}_3^{(1)} (x_i,\mu) | 0 \rangle_{\mathrm{SU(3)}} + \mathcal{T}_{3m,B}^{(V)} (x_i), \\
\langle s\bar{s}g | \mathcal{O}_3^{(1)} (x_i,\mu) | 0 \rangle &=& \langle s\bar{s}g | \mathcal{O}_3^{(1)} (x_i,\mu) | 0 \rangle_{\mathrm{SU(3)}} + \mathcal{T}_{3m,A+B}^{(V)} (x_i), \nonumber 
\end{eqnarray} 
where $x_i,~i=1,2,3$, are the momentum fractions for the quark, antiquark, and the gluon. $\mathcal{O}_3^{(1)} (x,\mu)$ is the Fourier transform of the nonlocal operator shown in Eq.~(\ref{3M}) that, with the constraint $\sum_i x_i =1$, can be written as 
\begin{equation}  
\mathcal{O}_3^{(1)} (x_i) = \bar{\xi}_n^q W \delta \Bigl(x_1 - \frac{\bP^{\dagger}}{\n\cdot p_M} \Bigr)\nn 
\Biggl[ \fms{\mathcal{B}}_{\perp} \delta \Bigl(x_3 - \frac{\bP^{\dagger}}{\n\cdot p_M}\Bigr) \Biggr] 
\gamma_5 W^{\dagger} \xi_n^{q'}. 
\end{equation}
As before, we relate the SU(3) limit matrix element in SU(3) to the pion LCDA, $\langle \mathcal{O}_3^{(1)} \rangle_{\mathrm{SU(3)}} = i (\n\cdot p_M)^2 f_{3\pi} \phi_{3\pi}$. 
As  with $\mathcal{T}_{m}^{(V)}$ and $\mathcal{T}_{m}^{(V,\pm)}$, the valence contributions $\mathcal{T}_{3m}^{(V)}$ respect the relations 
\begin{equation} 
\label{3mrela} 
\mathcal{T}_{3m,A+B}^{(V)} (x_i) = \mathcal{T}_{3m,A}^{(V)} (x_i) + \mathcal{T}_{3m,B}^{(V)} (x_i),~~~
\mathcal{T}_{3m,A}^{(V)} (x_1,x_2,x_3) = \mathcal{T}_{3m,B}^{(V)} (x_2,x_1,x_3)  
\end{equation} 
From Eqs.~(\ref{3su3}) and (\ref{3mrela}), we find 
\begin{equation} 
\label{3prel} 
f_{3\pi} \phi_{3\pi} (x_i) + 3 f_{3\eta} \phi_{3\eta} (x_i) = 
2f_{3K} \Bigl[\phi_{3K^-} (x_i) + \phi_{3K^+} (x_i) \Bigr] = 
2f_{3K} \Bigl[\phi_{3\bar{K}} (x_i) + \phi_{3K^0} (x_i) \Bigr].  
\end{equation} 

The three-particle LCDA defined in Eq.~(\ref{3pi}) does not have a convenient normalization.  We can redefine $\phi_{3M}$ so that  $\int [dx_i] \phi_{3M} (x_i) = 1$
by absorbing a scale-dependent factor into the decay constant $f_{3\pi} (\mu)$, 
\begin{eqnarray} 
\label{3pi2} 
\phi_{3\pi} (x_i,\mu) &=& 360 x_1 x_2 x_3^2 \Bigl[1 + \frac{\w_{3\pi}(\mu)}{2} (7x_3-3) + \cdots \Bigr], \\ 
\phi_{3\eta} (x_i,\mu) &=& 360 x_1 x_2 x_3^2 \Bigl[1 + \frac{\w_{3\eta} (\mu)}{2} (7x_3-3) + \cdots \Bigr],
\end{eqnarray} 
while $\phi_{3K,3\eta}$, due to SU(3) breaking corrections can be written as 
\begin{equation} 
\label{phi3k}
\phi_{3K^{\mp}} (x_i,\mu) = 360 x_1 x_2 x_3^2 \Bigl[1\pm \lambda_{3K} (\mu) (x_1 - x_2) + \frac{\w_{3K} (\mu)}{2} (7x_3-3) + \cdots \Bigr].
\end{equation} 
Comparing with Eq.~(\ref{3pi2}), $\lambda_{3K}$ gives an asymmetric distribution under exchange $x_1 \leftrightarrow x_2$ due to the SU(3) breaking between the quark and antiquark and is thus $\mathcal{O}(m_s/\Lambda)$. Due to the new normalization LCDAs, from Eq.~(\ref{3prel}) we easily see 
\begin{equation} 
\label{3preltt} 
f_{3\pi} + 3 f_{3\eta} = 4 f_{3K},~~~f_{3\pi}\w_{3\pi} + 3 f_{3\eta}\w_{3\eta} = 4 f_{3K}\w_{3K}.
\end{equation} 

The valence SU(3) breaking contributions for $f_{3K,3\eta} \phi_{3K,3\eta}$ arise in part from the mixing with the matrix element $\langle m_s \mathcal{O}^{(0)} \rangle$ under renormalization. For example, the mixing for $K^-$ can be calculated through the Feynman diagram Fig.~\ref{conmix}-(b), and we find the renormalization group equation
\begin{equation} 
\mu \frac{d}{d\mu} \Bigl[f_{3K} \phi_{3K^-} (x_i, \mu) \Bigr]_{\mathrm{mixed}} 
= \frac{\alpha_s C_F}{2\pi} m_s f_K \frac{x_3^2}{(x_1+x_3)^2} \phi_{K^-} (x_1+x_3). 
\end{equation} 
Using the orthogonality to each term in Eq.~(\ref{phi3k}) under the integration $\int [dx_i]$, 
we obtain the mixed renormalization group equations for $f_{3K},~\lambda_{3K},$ and $\w_{3K}$,
\begin{eqnarray} 
\mu \frac{d}{d\mu} f_{3K} &=& -\gamma_{3f} f_{3K} + \frac{\alpha_s C_F}{\pi} m_s f_K \Bigl(\frac{1}{12} + \frac{1}{20} a_1^K \Bigr),  \\ 
\mu \frac{d}{d\mu} (f_{3K} \lambda_{3K} ) &=& -\gamma_{3\lambda} f_{3K} \lambda_{3K} + \frac{\alpha_s C_F}{4\pi} m_s f_K \Bigl(-\frac{7}{6} + \frac{7}{10} a_1^K + a_2^K\Bigr), \\
\mu \frac{d}{d\mu} (f_{3K} \w_{3K} ) &=& -\gamma_{3\w} f_{3K} \w_{3K} + \frac{\alpha_s C_F}{4\pi} m_s f_K \Bigl(\frac{3}{45} + \frac{3}{5} a_1^K + \frac{6}{15} a_2^K\Bigr), 
\end{eqnarray} 
where we included the only lowest two coefficients $a_{1,2}^K$ in $\phi_{K^-}$ for the simplicity, and $\gamma_{\{3,3\lambda,3\w\}}$ are the anomalous dimesions for $\{f_{3M},~f_{3M}\lambda_{3M},~ f_{3M}\w_{3M}\}$ in the massless limit \cite{gamma}
\begin{equation} 
\gamma_{3f} = \frac{\alpha_s}{4\pi} \frac{110}{9},~~~\gamma_{3\lambda} = \frac{\alpha_s}{4\pi} \frac{139}{9},~~~\gamma_{3\w} = \frac{\alpha_s}{4\pi} \frac{208}{9}. 
\end{equation} 
The anomalous dimensions for the quark mass is $\gamma_m = 8 \alpha_s/(4\pi)$ and $\gamma_n$ for $a_n^M$ are given in Eq.~(\ref{gamman}). Combining, we obtain the  solutions for the renormalization group equations, 
\begin{eqnarray} 
f_{3K} (\mu) &=& \LL^{\frac{55}{9\beta_0}} f_{3K} (\mu_0)+ \frac{2}{19} \Biggl[\LL^{\frac{4}{\beta_0}} - \LL^{\frac{55}{9\beta_0}}\Biggr] (m_s f_K) (\mu_0) \nonumber \\ 
&&+ \frac{6}{65}  \Biggl[\LL^{\frac{55}{9\beta_0}}-\LL^{\frac{68}{9\beta_0}}\Biggr] (m_s a_1^K f_{K}) (\mu_0), \\ 
(f_{3K} \lambda_{3K}) (\mu) &=& \LL^{\frac{139}{18\beta_0}} (f_{3K} \lambda_{3K}) (\mu_0)+ \frac{14}{67} \Biggl[\LL^{\frac{139}{18\beta_0}} - \LL^{\frac{4}{\beta_0}}\Biggr] (m_s f_K) (\mu_0) \nonumber \\ &&- \frac{14}{5}  \Biggl[\LL^{\frac{139}{18\beta_0}}-\LL^{\frac{68}{9\beta_0}}\Biggr] (m_s a_1^K f_{K}) (\mu_0)  \nonumber \\
&&+ \frac{4}{11}  \Biggl[\LL^{\frac{139}{18\beta_0}}-\LL^{\frac{86}{9\beta_0}}\Biggr] (m_s a_2^K f_{K}) (\mu_0) \\ 
(f_{3K} \w_{3K}) (\mu) &=& \LL^{\frac{104}{9\beta_0}} (f_{3K} \w_{3K}) (\mu_0)+ \frac{1}{170} \Biggl[\LL^{\frac{4}{\beta_0}} - \LL^{\frac{104}{9\beta_0}}\Biggr] (m_s f_K) (\mu_0) \nonumber \\ &&+ \frac{1}{10}  \Biggl[\LL^{\frac{68}{9\beta_0}}-\LL^{\frac{104}{9\beta_0}}\Biggr] (m_s a_1^K f_{K}) (\mu_0)  \nonumber \\
&&+\frac{2}{15}  \Biggl[\LL^{\frac{86}{9\beta_0}}-\LL^{\frac{104}{9\beta_0}}\Biggr] (m_s a_2^K f_{K}) (\mu_0),
\end{eqnarray} 
which agree with the full QCD result \cite{Ball:2006wn}. 

\section{Exact relations between twist-3 LCDAs} 
The relations between twist-3 LCDAs including light quark masses have been studied in full QCD \cite{Ball:2006wn,Ball:1998je}. In this section, we consider the relations again using SCET. As a result we express $\phi_{\pm}^K$ in terms of the quark masses, $a_n^K$, and $\{f_{3K}, \lambda_{3K}, \w_{3K}\}$, equivalent to what was found in Ref.~\cite{Ball:2006wn}. Further we consider $\phi^{\eta}_{\pm}$ in order to confirm the relations between the mesons $\pi,~K$, and $\eta$ shown in Eqs.~(\ref{pmrel}) and (\ref{psrel}). 

Keeping the light quark masses in, for example, the $K^-$ system, Eq.~(\ref{relation}) is modified to 
\begin{eqnarray}  
\label{rel1} 
&&\frac{R_K}{2} \Bigl[ \bu\phi_+^{K^-} (u) - u\phi_-^{K^-} (u) + \frac{m_s-m_u}{\mu_K} \phi_{K^-} (u)\Bigr] \\
&&=  \Biggl[\int^{\bu}_0 db \frac{\phi_{3K} (u,\bu-b,b)}{b} 
-\int^{u}_0 db \frac{\phi_{3K} (u-b,\bu,b)}{b} \Biggr], \nonumber 
\end{eqnarray} 
where $R_K = f_K \mu_K/f_{3K}$ and  $\phi_{3K}$ has been defined in Eq.~(\ref{phi3k}). The term involving quark masses come from the matrix elements of $\mathcal{K}_{\pm}$ shown in Eqs.~(\ref{o+1}) and (\ref{o-1}). We need one more relation in order to solve for $\phi_{\pm}^{K^-}$ simultaneously.  For the pion, we had the isospin symmetry relation, Eq.~({\ref{isorel}), which is unavailable for the $K^-$.

Instead, we can obtain another relation starting from Eq.~(\ref{phis}), keeping exact power-counting in $\lambda$.
Applying $(\n^{\nu}/2) (-i\partial_x^{\mu}+i\partial_y^{\mu})$ to both sides of Eq.~(\ref{phis}), we have 
\begin{eqnarray} 
&&\frac{\n^{\nu}}{2} (-i\partial_x^{\mu}+i\partial_y^{\mu}) \langle K^- | \bar{q}^s (x) \sigma_{\mu\nu} \gamma_5 [x, y] q^u (y) | 0 \rangle \nonumber \\ 
\label{sigma1}
&&=-f_K \mu_K \n\cdot p_K \int^1_0 du e^{i\n\cdot p_K (ux + \bu y)} \Bigl[\frac{\phi_{\sigma}(u)}{3} - \frac{2u-1}{12} \frac{\partial \phi_{\sigma}}{\partial u} (u) \Bigr],
\end{eqnarray} 
where $q = \xi_n + \xi_{\n}$, and we have used the relations 
\begin{eqnarray} 
(-i\partial_x^{\mu}+i\partial_y^{\mu}) \Delta^{\nu} &=& -2i g^{\mu\nu}, \\
(-i\partial_x^{\mu}+i\partial_y^{\mu}) e^{i\n\cdot p_K (ux + \bu y)} &=& \frac{n^{\mu}}{2} \n\cdot p_K (2u-1) e^{i\n\cdot p_K (ux + \bu y)}. 
\end{eqnarray} 
As before, we have set all the coordinates to be on the same lightcone, $x^{\mu} = x\n^{\mu},~ y^{\mu} = y\n^{\mu}$, and thus $\partial_x^{\mu} f(x) = \partial_x^{\mu} f(x)\Bigl|_{x_{\perp},n\cdot x = 0}$.
We can write the left-hand side of Eq.~(\ref{sigma1}) as 
\begin{eqnarray} 
&&\frac{\n^{\nu}}{2}(-i\partial_x^{\mu}+i\partial_y^{\mu}) \langle \bar{q}^s (x) \sigma_{\mu\nu} \gamma_5 [x, y] q^u (y) \rangle = \frac{\n^{\nu}}{2}\langle \bar{q}^s W(x) \Bigl(\mP^{\dagger\mu} + \mP^{\mu} \Bigr) \sigma_{\mu\nu} \gamma_5 W^{\dagger} q^u (y) \rangle \nonumber \\
\label{sigmal1} 
&&~~~~~~~+\frac{\n^{\nu}}{2}\frac{2}{x-y} \langle \bar{q}^s W(x) \int^x_y dz \Bigl[W^{\dagger} iD^{\mu} W\Bigr] (z) \sigma_{\mu\nu} \gamma_5 W^{\dagger} q^u (y) \rangle, 
\end{eqnarray} 
using the identity for the Wilson link $[x,y]$
\begin{eqnarray} 
(-i\partial_x^{\mu}+i\partial_y^{\mu}) [x,y] &=& -i\partial_x^{\mu} W(x) W^{\dagger}(y) + W(x)~i\partial_y^{\mu} W^{\dagger} (y) \nonumber \\
\label{iden2} 
&&+\frac{2}{x-y} W(x) \int^x_y dz \Bigl[W^{\dagger} iD^{\mu} W\Bigr] (z) W^{\dagger} (y). 
\end{eqnarray} 

Applying Eq.~(\ref{iden1}) and the equation of motion $i\fmsl{D} ~q (x) = m q (x) $ to the first term on the right-hand side of Eq.~(\ref{sigmal1}), we have 
\begin{eqnarray} 
\bar{q}^s W(x) \mP^{\dagger \mu} \sigma_{\mu\nu} \gamma_5 W^{\dagger} q^u (y) 
&=& -i \bar{q}^s W(x) \mP^{\dagger}_{\nu} \gamma_5 W^{\dagger} q^u (y) + im_s \bar{q}^s W(x) \gamma_{\nu} \gamma_5 W^{\dagger} q^u (y) \nonumber \\
\label{sigmal21}
&&- i\bar{q}^s W \Bigl[ W^{\dagger} i\overleftarrow{\fmsl{D}_{\perp}} W\Bigr] (x) \gamma_{\nu} \gamma_5 W^{\dagger}q^u (y),  \\
\bar{q}^s W(x) \mP^{\mu} \sigma_{\mu\nu} \gamma_5 W^{\dagger} q^u (y) 
&=& -i \bar{q}^s W(x) \mP_{\nu} \gamma_5 W^{\dagger} q^u (y) + im_u \bar{q}^s W(x) \gamma_{\nu} \gamma_5 W^{\dagger} q^u (y) \nonumber \\
\label{sigmal22} 
&&- i\bar{q}^s W (x) \gamma_{\nu} \gamma_5 \Bigl[W^{\dagger} i\fmsl{D}_{\perp} W\Bigr] (x)  W^{\dagger}q^u (y). 
\end{eqnarray} 
Combining Eqs.~(\ref{sigmal1}), (\ref{sigmal21}), and (\ref{sigmal22}), to lowest order in $\lambda$, we can rewrite the left-hand side of Eq.~(\ref{sigma1}) as 
\begin{eqnarray} 
\label{sigmal2} 
&&\frac{\n^{\nu}}{2} (-i\partial_x^{\mu}+i\partial_y^{\mu}) \langle \bar{q}^s (x) \sigma_{\mu\nu} \gamma_5 [x, y] q^u (y) \rangle =  \\ 
&& -\frac{i\n\cdot p_K}{2} \Bigl\langle  \bar{\xi}_n^s W (x) \gamma_5 W^{\dagger} \xi_{\n}^u (y) + \bar{\xi}_{\n}^s W (x) \gamma_5 W^{\dagger} \xi_{n}^u (y) \Bigr\rangle
+i(m_s+m_u) \Bigl\langle  \bar{\xi}_n^s W (x) \nn \gamma_5 W^{\dagger} \xi_{n}^u (y) \Bigr\rangle \nonumber \\ 
&&+\frac{1}{x-y} \Bigl\langle  \bar{\xi}_n^s W (x) \nn \int^x_y dz [2z-(x+y)] \fms{\mathcal{B}}_{\perp} (z) \gamma_5 W^{\dagger} \xi_n^u (y) \Bigr\rangle.  \nonumber 
\end{eqnarray} 
Applying Eqs.~(\ref{phi+}), (\ref{phi-}), and (\ref{3M}) to the above equation, we find 
\begin{eqnarray} 
\label{sigmal3}
&&\frac{\n^{\nu}}{2} (-i\partial_x^{\mu}+i\partial_y^{\mu}) \langle K^- |\bar{q}^s (x) \sigma_{\mu\nu} \gamma_5 [x, y] q^u (y) | 0 \rangle =  \\ 
&&+\frac{1}{2} f_K \n\cdot p_K (m_s + m_u) \int^1_0 e^{i\n\cdot p_K (ux + \bu y)} \phi_{K^-} (u) \nonumber \\
&&-\frac{1}{4} f_K \mu_K \n\cdot p_K \int^1_0 e^{i\n\cdot p_K (ux + \bu y)} \Bigl[\phi_+^{K^-} (u) + \phi_-^{K^-} (u) \Bigr] \nonumber \\
&&+\frac{f_{3K} \n\cdot p_K}{x-y} \int^x_y dz \int^1_0 [du_i] e^{i\n\cdot p_K (u_1x + u_2 y+ u_3 z)}
K(u_1,u_2,u_3), \nonumber 
\end{eqnarray}
where $K(u_i)$ in the last line is
\begin{equation} 
K (u_1, u_2, u_3) = \Biggl(\frac{\partial}{\partial u_1}+\frac{\partial}{\partial u_2}-2\frac{\partial}{\partial u_3} \Biggr) \phi_{3K} (u_1, u_2, u_3).
\end{equation}
Comparing Eq.~(\ref{sigmal3}) to the right-hand side of Eq.~(\ref{sigma1}), finally we obtain  
\begin{eqnarray} 
\label{rel2} 
&&\int^u_0 \frac{db}{b} K (u-b, 1-u, b) - \int^{\bu}_0 \frac{db}{b} K (u, 1-u-b, b) \\
&&=R_K \frac{\partial}{\partial u} \Biggl[\frac{\phi_{\sigma}^{K^-}}{3} - \frac{2u-1}{12} \frac{\partial}{\partial u} \phi_{\sigma}^{K^-} - \frac{1}{4} \Bigl(\phi_+^{K^-} (u) + \phi_-^{K^-} (u) \Bigr) + \frac{m_s+m_u}{2\mu_K} \phi_{K^-} (u) \Biggr], \nonumber \\
&&=R_K \Biggl[\frac{m_s+m_u}{2\mu_K} \frac{\partial}{\partial u} \phi_{K^-} (u) - \frac{\partial}{2\partial u} \Bigl(\bu \phi_+^{K^-} (u) + u \phi_-^{K^-} (u) \Bigr) - \Bigl(\phi_+^{K^-} (u) - \phi_-^{K^-} (u) \Bigr) \Biggr], \nonumber 
\end{eqnarray}
which is our second relation needed to determine $\phi_\pm^{K^-}$.

From Eqs.~(\ref{rel1}) and (\ref{rel2}), we can determine $\phi_{\pm}^{K^-}$. To begin with, expand $\phi_{\pm}^{K^-}$ in terms of Jacobi polynomials $P_n^{(0,1)}$ and $P_n^{(1,0)}$,
\begin{eqnarray} 
\label{+form} 
\phi_{+}^{K^-} (u) &=& 2u \Biggl[1+ \sum_{n=1}^{\infty} \Bigl(\alpha_{3/2+n} + \beta_{3/2+n} \Bigr) 
P_n^{(0,1)} (2u-1) \\
&&+ \sum_{n=1}^{\infty} \Bigl(\alpha_{3/2+n}^G + \beta_{3/2+n}^G \Bigr) P_n^{(0,1)} (2u-1) + F_+ (u) \Biggr], \nonumber \\
\label{-form} 
\phi_{-}^{K^-} (u) &=& 2\bu \Biggl[1+ \sum_{n=1}^{\infty} (-1)^n \Bigl(\alpha_{3/2+n} - \beta_{3/2+n} \Bigr) 
P_n^{(1,0)} (2u-1) \\
&&+ \sum_{n=1}^{\infty} (-1)^n \Bigl(\alpha_{3/2+n}^G - \beta_{3/2+n}^G \Bigr) P_n^{(1,0)} (2u-1) + F_- (u) \Biggr], \nonumber 
\end{eqnarray} 
where $\alpha_{3/2+n}^{(G)}$ $(\beta_{3/2+n}^{(G)})$ are (anti-)symmetric coefficients under exchange $u\leftrightarrow \bu$, and the superscript $(G)$ denotes coefficients coming from the three-particle LCDA $\phi_{3K^-}$. 
$F_{\pm}$ are SU(3) breaking in the asymptotic form, obtained by neglecting the contributions from $\phi_{3K^-}$.  In this case, Eqs.~(\ref{rel1}) and (\ref{rel2}) simplify to 
\begin{eqnarray} 
\label{rela1} 
&&\bu \phi_{+}^{K^-}  - u \phi_{-}^{K^-}  + \ka_- \phi_{K^-} = 0, \\ 
\label{rela2} 
&&\phi_{+}^{K^-} - \phi_{-}^{K^-} + \bu \frac{\partial \phi_+^{K^-}}{\partial u} + u \frac{\partial \phi_-^{K^-}}{\partial u} - \ka_+ \frac{\partial \phi_{K^-}}{\partial u} = 0, 
\end{eqnarray} 
where $\ka_{\pm} = (m_s\pm m_u)/\mu_K$. 
Keeping only the lowest two coefficients  $a_{1,2}^{K}$ in $\phi_{K^-}$ and choosing the normalization $\int du\, \phi_p^{K^-} (u) = 1$, we find  
\begin{equation} 
\label{fpm} 
F_{\pm} (u) = \frac{3}{2} b_+ \Bigl(2 + \ln u\bu\Bigr) + \frac{3}{2} b_- \Bigl(\mp1+ \ln \frac{u}{\bu}\Bigr),  
\end{equation} 
where $b_{\pm} = \ka_{\pm} - 3\ka_{\mp} a_1^K + 6 \ka_{\pm} a_2^K$. 
The coefficients $\alpha_{3/2+n},~\beta_{3/2+n}$ can be be obtained from Eqs.~(\ref{rela1}) and (\ref{rela2}),  
\begin{eqnarray} 
&&\alpha_{5/2} = - 3\ka_- a_1^K + 6 \ka_+ a_2^K,~~~ \beta_{5/2} = 9\ka_- a_1^K - 18 \ka_+ a_2^K, \\
&&\alpha_{7/2} = 9 \ka_+ a_2^K,~~~ \beta_{7/2} = -\frac{9}{2} \ka_- a_2^K. \nonumber 
\end{eqnarray} 
Then, including the contributions from $\phi_{3K^-}$ in Eqs.~(\ref{rel1}) and (\ref{rel2}), $\alpha^{G}_{3/2+n},~\beta^{G}_{3/2+n}$ are 
\begin{eqnarray} 
&&\alpha_{5/2}^G = \beta_{5/2}^G = 0,~~~\alpha_{7/2}^G = \frac{30}{R_K},~~~\beta_{7/2}^G = 0, \\ 
&&\alpha_{9/2}^G = -3\frac{\w_{3K}}{R_K},~~~\beta_{9/2}^G = 10 \frac{\lambda_{3K}}{R_K}. \nonumber 
\end{eqnarray} 

Using Eq.~(\ref{3rel}), we finally find 
\begin{eqnarray} 
\label{phipk-} 
\phi_p^{K^-} (u) &=& 1+3(\ka_+-3\ka_-a_1^K + 6\ka_+a_2^K) +\frac{3}{2} (\ka_++\ka_-) (1-3a_1^K+ 6a_2^K) \ln u \\ 
&&+ \frac{3}{2} (\ka_+ - \ka_-) (1+3a_1^K+ 6a_2^K) \ln \bu 
+C_1^{1/2}(2u-1) \Bigl(-\frac{3}{2}\ka_- + \frac{27}{2} \ka_+ a_1^K -27\ka_- a_2^K\Bigr) \nonumber \\
&&+C_2^{1/2}(2u-1) \Bigl(\frac{30}{R_K} - 3\ka_-a_1^K + 15\ka_+ a_2^K\Bigr) 
+ C_3^{1/2}(2u-1) \Bigl(10 \frac{\lambda_{3K}}{R_K} - \frac{9}{2} \ka_-a_2^K\Bigr) \nonumber \\
&&-3 C_4^{1/2}(2u-1) \frac{\w_{3K}}{R_K}+ \cdots, \nonumber \\
\label{phisk-} 
\phi_{\sigma}^{K^-} (u) &=& 6u\bu \Biggl[1+\frac{3}{2} \ka_+ -\frac{15}{2} \ka_-a_1^K +15 \ka_+ a_2^K
+\frac{3}{2} (\ka_++\ka_-) (1-3a_1^K+ 6a_2^K) \ln u \\ 
&&+ \frac{3}{2} (\ka_+ - \ka_-) (1+3a_1^K+ 6a_2^K) \ln \bu 
+C_1^{3/2}(2u-1) \Bigl(3\ka_+ a_1^K -\frac{15}{2} \ka_-a_2^K\Bigr) \nonumber \\
&&+C_2^{3/2}(2u-1) \Bigl(\frac{5}{R_K} -\frac{\w_{3K}}{2R_K} + \frac{3}{2} \ka_+ a_2^K \Bigr) 
+ C_3^{3/2}(2u-1) \frac{\lambda_{3K}}{R_K} + \cdots \Biggr]. \nonumber 
\end{eqnarray} 
We can easily confirm that $\phi_{p,\sigma}^{K^+} (u) = \phi_{p,\sigma}^{K^-} (\bu)$ and $\phi_{\pm}^{K^+} (u) = \phi_{\mp}^{K^-} (\bu)$ by replacing $\{\ka_-,a_1^K,\lambda_{3K}\} \to -\{\ka_-,a_1^K,\lambda_{3K}\}$ in Eqs.~(\ref{+form}), (\ref{-form}), (\ref{phipk-}), and (\ref{phisk-}). 

In a similar fashion, we can obtain $\phi_{p,\sigma}^{\eta}$,
\begin{eqnarray} 
\label{phipe} 
\phi_p^{\eta} (u) &=& 1+ 3\ka_+^{\eta}( 1+ 6 a_2^{s\bs}) 
+\frac{3}{2} \ka_+^{\eta} (1+ 6a_2^{s\bs}) \ln u\bu \\ 
&&+C_2^{1/2}(2u-1) \Bigl(\frac{30}{R_{\eta}} + 15 \ka_+ a_2^{s\bs} \Bigr) 
-3 C_4^{1/2}(2u-1) \frac{\w_{3\eta}}{R_{\eta}}+ \cdots, \nonumber \\
\label{phise} 
\phi_{\sigma}^{\eta} (u) &=& 6u\bu \Biggl[1 +\frac{3}{2} \ka_+^{\eta} (1+ 10 a_s^{s\bs}) 
+\frac{3}{2} \ka_+^{\eta} (1+ 6a_2^{s\bs}) \ln u\bu \\ 
&&+C_2^{3/2}(2u-1) \Bigl(\frac{5}{R_{\eta}} -\frac{\w_{3\eta}}{2R_{\eta}} + \frac{3}{2} \ka_+^{\eta} a_2^{s\bs} \Bigr) + \cdots \Biggr], \nonumber 
\end{eqnarray} 
where $\ka_+^{\eta} = 4f_{s\bs}m_s/(3f_{\eta}\mu_{\eta}) \sim 4\ka_+/3$ at leading SU(3) breaking, with $\mu_{\eta} = 3m^2_{\eta}/(m_u+m_d+4m_s) \sim 3m^2_{\eta}/(4m_s)$, and $a_n^{s\bs}$ comes from the leading matrix element
\begin{equation} 
\langle s\bs | \bar{\xi}_n^s W \delta\Bigl(u-\frac{\bP^{\dagger}}{\n\cdot p_{\eta}}\Bigr) \nn \gamma_5 W^{\dagger} \xi_n^s | 0 \rangle = -if_{s\bs} \frac{\n\cdot p_{\eta}}{2} 6u\bu \Bigl[1+ a_2^{s\bs} C_2^{3/2} (2u-1) + \cdots \Bigr].  
\end{equation}

From the above results, we can check the relation Eq.~(\ref{psrel}) neglecting $\mathcal{O}(m_s^2/\Lambda^2)$.  In this case, we take $\ka_+ = \ka_-,~ m_s\{f_{K,\eta}, f_{3K,3\eta}, \mu_{K,\eta}, a_2^{K,s\bs} \} = m_s \{f_{\pi},f_{3\pi}, \mu_{\pi}, a_2^{\pi}\}$, and $m_s\{a_1^K, \lambda_{3K} \} = 0$. 
As an example, for $\phi_p^M$, we have 
\begin{eqnarray} 
f_{\pi}\mu_{\pi} \phi_p^{\pi} + 3f_{\eta}\mu_{\eta} \phi_p^{\eta} &=& f_{\pi}\mu_{\pi}  + 3f_{\eta}\mu_{\eta} + f_{\pi} \Bigl[12m_s + 72 m_s a_2^{\pi} + 6m_s(1+6a_2^{\pi}) \ln u\bu \Bigr] \nonumber \\
&&+C_2^{1/2}(2u-1) \Bigl[30f_{3\pi} + 90 f_{3\eta} + 60 m_s f_{\pi} a_2^{\pi} \Bigr] \nonumber \\
\label{p+e} 
&&-C_4^{1/2}(2u-1) \Bigl[3f_{3\pi}\w_{3\pi} + 9f_{3\eta}\w_{3\eta}  \Bigr] + \cdots, \\ 
2f_K \mu_K \Bigl[\phi_p^{K^+} + \phi_p^{K^-} \Bigr] &=& 4f_K\mu_K + 12f_{\pi}(m_s + 6 m_s a_2^{\pi} )
+ 6m_s f_{\pi} (1+6a_2^{\pi}) \ln u\bu \nonumber \\
&&+C_2^{1/2}(2u-1) \Bigl[120 f_{3K} +60 m_s f_{\pi} a_2^{\pi} \Bigr] \nonumber \\
\label{k+k} 
&&-12 C_4^{1/2}(2u-1) 3f_{3K}\w_{3K}  + \cdots. 
\end{eqnarray} 
Using Eq.~(\ref{3preltt}) in the above equations, we confirm the SU(3) relation for $\phi_p^M$.  We can similarly check that the relations for $\phi_{\pm}^M$ and $\phi_{\sigma}^M$ are satisfied.

\section{Conclusion} 
In this paper we have studied the lightcone formalism for light pseudoscalar mesons using SCET. The lightcone conformal symmetry is transparent within the framework of SCET, with the twist expansion corresponding to the SCET power-counting expansion in $\lambda$.  The LCDAs are well-defined in a gauge-invariant way at any given order in $\lambda$.  
Relations between LCDAs, equivalent to established full QCD results, have been rederived in an independent way, not using the term-by-term matching between full QCD and SCET as suggested in Ref.~\cite{Hardmeier:2003ig}.

We also investigated leading SU(3) breaking corrections to the lightcone formalism. The leading SU(3) breaking effects in LCDAs can be realized in SCET as the time-ordered products of nonlocal operators and $\mathcal{L}_{m_s}^{(1)}$, the mass term in the SCET Lagrangian.  The contributions can be categorized as ``sea" and ``valence", with the former independent on the type of light meson.  Thus, the SU(3) breaking effects for each meson can be specified solely by the valence contribution. Analyzing these, we obtain leading SU(3) breaking relations between the light meson $\pi,K,\eta$ at leading and subleading order in $\lambda$, given in Eqs.~(\ref{su3rel}) and (\ref{psrel}). 
The leading and subleading relations Eqs.~(\ref{su3rel}) and (\ref{psrel}) imply the SU(3) breaking relations between the decay constants $f_{\pi} + 3 f_{\eta} = 4f_K$ and $\mu_{\pi} + 3\mu_{\eta} = 4\mu_K$.
Applying these results to the original relations, we easily show    
\begin{eqnarray} 
\phi_{\pi} (x) + 3\phi_{\eta} (x) &=& 2 \Bigl[\phi_{K^-} (x) + \phi_{K^+} (x) \Bigr] = 2 \Bigl[\phi_{\bar{K}^0} (x) + \phi_{K^0} (x) \Bigr], \\
\phi_{\pm, p,\sigma}^{\pi} (x) + 3 \phi_{\pm, p,\sigma}^{\eta} (x) &=& 2
\Bigl[\phi_{\pm, p,\sigma}^{K^-} (x) + \phi_{\pm, p,\sigma}^{K^+} (x) \Bigr] = 
\Bigl[\phi_{\pm, p,\sigma}^{\bar{K}^0} (x) + \phi_{\pm, p,\sigma}^{K^0} (x) \Bigr]. 
\end{eqnarray} 
which were obtained in Refs.~\cite{Chen:2003fp,Chen:2005js} by  applying ChPT to the lightcone formalism. 

The results  presented here could be extended to light vector mesons and can be applied to hard processes with energetic particles.  For example, our results could contribute to the clarification of the difference between $B\to \pi\pi$ and $B \to \pi K$ under the factorization approach. In $\eta-\eta'$ systems, the mixing angle could be extracted more accurately allowing for SU(3) breaking.  More studies in this direction may give significant corrections to present theoretical results, and hence allow for better interpretation of experimental data or might even help us identify new physics. 

\section*{Acknowledgments}
C.~K.~were supported in part by the Department of Energy under grant numbers DE-FG02-05ER41368 and DE-FG02-05ER41376.  A.~K.~L.~is supported in part by the National Science Foundation under Grant No.~PHY-0546143 and in part  by the Research Corporation.

\end{document}